\documentclass[preprint,12pt,authoryear]{elsarticle}

\usepackage{amssymb}
\usepackage{amsmath}

\usepackage{hyperref}

\usepackage[htt]{hyphenat}

\usepackage{subcaption}

\usepackage{pdflscape}
\usepackage{listings}
\usepackage{xcolor}
\usepackage{subcaption} 

\usepackage{svg}

\let\cite\citep

\journal{Astronomy and Computing}

\begin{document}

\begin{frontmatter}



\title{iDaVIE v1.0: A Virtual Reality Tool for Interactive Analysis of Astronomical Data Cubes}


\affiliation[idia]{organization={Inter-University Institute for Data Intensive Astronomy, University of Cape Town},
      addressline={Private Bag X3}, 
      city={Rondebosch},
      postcode={7701}, 
      country={South Africa}}

\affiliation[uct]{organization={Department of Astronomy, University of Cape Town},
    addressline={Private Bag X3}, 
    city={Rondebosch},
    postcode={7701}, 
    country={South Africa}}

\affiliation[armagh]{organization={Armagh Observatory and Planetarium},
    addressline={College Hill}, 
    city={Armagh},
    postcode={BT61 9DB}, 
    country={Northern Ireland}}

\affiliation[inaf]{organization={Astrophysical Observatory of Catania, National Institute for Astrophysics},
    addressline={Via Santa Sofia 78}, 
    city={Catania},
    postcode={95123}, 
    country={Italy}}

\affiliation[ira]{organization={INAF - Istituto di Radioastronomia},
    addressline={via Gobetti 101}, 
    city={Bologna},
    postcode={40129}, 
    country={Italy}}
    
\affiliation[kapetyn]{organization={Kapteyn Astronomical Institute, University of Groningen},
    addressline={Landleven 12}, 
    city={Groningen},
    postcode={9747AD}, 
    country={The Netherlands}}

\affiliation[oac]{organization={INAF - Osservatorio Astronomico di Cagliari},
    addressline={Via della Scienza 5}, 
    city={Selargius},
    postcode={09047}, 
    country={Italy}}

\affiliation[uwc_phys]{organization={Department of Physics and Astronomy, University of the Western Cape},
    addressline={Robert Sobukwe Road}, 
    city={Bellville},
    postcode={7535}, 
    country={South Africa}}
    
\affiliation[uct_phys]{organization={Department of Physics, University of Cape Town},
    addressline={Private Bag X3}, 
    city={Rondebosch},
    postcode={7701}, 
    country={South Africa}}

\author[idia,uct,armagh]{Alexander Sivitilli\corref{cor1}} 

\cortext[cor1]{Corresponding author}

\ead{alexander.sivitilli@armagh.ac.uk}

\author[uct,idia,ira]{Lucia Marchetti} 

\author[idia]{Angus Comrie} 

\author[idia,inaf]{P. Cilliers Pretorius} 

\author[kapetyn]{Thijs (J.M.) van der Hulst} 

\author[inaf]{Fabio Vitello} 

\author[uct]{D.J. Pisano} 

\author[inaf]{Ugo Becciani} 

\author[idia,uwc_phys]{A. Russell Taylor} 

\author[oac]{Paolo Serra}

\author[uct_phys]{Mayhew Steyn}

\author[idia]{Michaela van Zyl}

\begin{abstract}
As modern astronomy confronts unprecedented data volumes, automated pipelines and machine-learning techniques have become essential for processing and analysis. As these workflows grow more complex, astronomers also require input and inspection tools that can keep pace. To address challenges in navigating multidimensional datasets for quality control and scientific interpretation, we present the immersive Data Visualisation Interactive Explorer (iDaVIE), a virtual reality (VR) software suite developed in collaboration with the astronomy community.

iDaVIE enables users to import and render large 3D data cubes within a VR environment, offering real-time tools for selection, cropping, catalogue overlays, and exporting results back into existing pipelines. Built on the Unity engine and SteamVR, the system uses custom plug-ins for efficient data parsing, downsampling, and statistical calculations.

The software has already been integrated into workflows such as verifying HI data cubes from MeerKAT, ASKAP, and APERTIF, refining detection masks, and identifying new sources. Its intuitive interface aims to reduce the cognitive load associated with higher-dimensional data, allowing researchers to focus more directly on scientific goals.

As an open-source, scalable, and adaptable platform, iDaVIE supports continued development and integration with other tools. Version 1.0 marks a significant milestone, with planned enhancements including subcube loading, advanced rendering modes, video-generation scripts, and collaborative capabilities. By pairing immersive visualisation with robust interaction tools, iDaVIE seeks to transform how researchers engage with complex datasets and enhance productivity in the era of big data.

\end{abstract}

\begin{graphicalabstract}
\includegraphics[width=\textwidth]{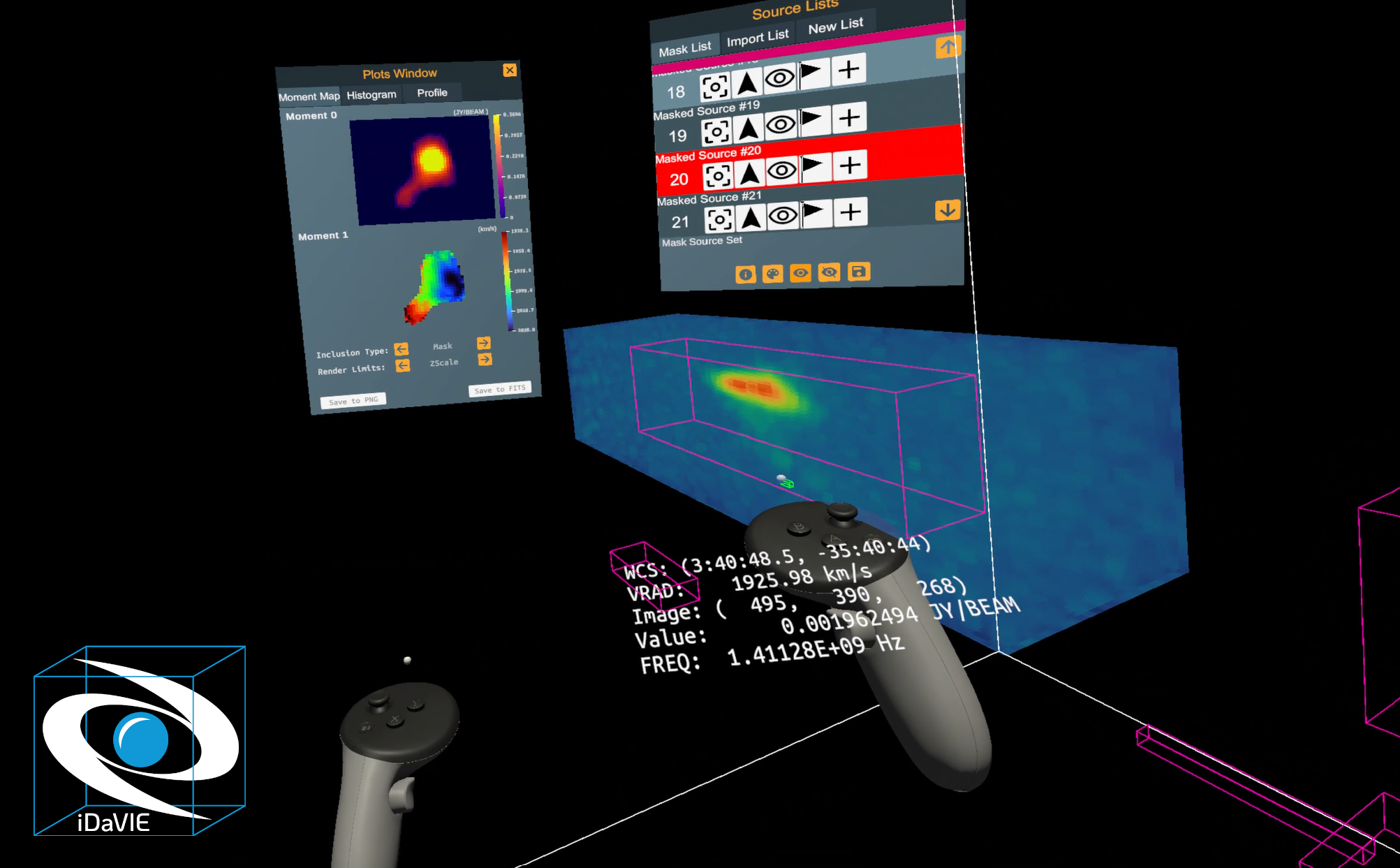}
\end{graphicalabstract}

\begin{highlights}
\item Immersive VR tool designed to support intuitive exploration of large 3D astronomy datasets
\item Real-time selection, masking, and quantitative analysis for data cubes
\item Integrates Unity and SteamVR with custom native plug-ins for efficient FITS data handling
\item Featured in MeerKAT, ASKAP, and APERTIF studies for source verification and mask refinement
\item Open-source and adaptable for future collaborative and multidisciplinary research
\end{highlights}

\begin{keyword}
big-data \sep virtual-reality \sep visualisation \sep immersive-visualisation \sep radio-astronomy \sep 3D-data \sep astronomical-software


\end{keyword}

\end{frontmatter}

\noindent{\small\textit{This manuscript has been accepted for publication in Astronomy \& Computing.
The iDaVIE software is available at \url{https://github.com/idia-astro/iDaVIE} and archived on Zenodo
(\href{https://doi.org/10.5281/zenodo.4614115}{DOI: 10.5281/zenodo.4614115}).}}



\section{Introduction}
\label{intro}

Astronomy, like many other sciences, is currently undergoing a radical shift in the volumes of digital data that it captures and analyses. Increasing sensitivity and resolution of telescopes, along with the refinement of techniques such as interferometry, have led to a surge in the amount of data generated by astronomical observations. This has resulted in the emergence of big data in astronomy, where datasets for projects can reach terabytes or even petabytes in size. For instance, the Vera C. Rubin Observatory is expected to produce up to 20 terabytes of data per night \cite{lsst_paper}. These advancements allow astronomers to capture more detailed and comprehensive data about the universe, but they also present significant challenges in terms of data storage, processing, and analysis. The ability to efficiently manage and extract meaningful insights from these vast datasets is crucial for advancing our understanding of the cosmos \cite{bigdata_paper}.

 This is especially the case with radio astronomy, given that the Square Kilometre Array (SKA) is expected to go live at the end of the decade. The SKA will be combining the celestial radio signals from thousands of separate antennas, producing up to 4.1 Petabits of data per second \cite{ska_data_paper}. This involves not only the appropriate architecture to collect and process the raw data, but also the automated pipelines to reduce it to manageable formats and media. Exploring efficient methods of interfacing with these large datasets directly, along with the results of their analysis, becomes increasingly necessary to refine and guide the reduction process from beginning to end. In fact, multiple stages of this process require the ability to efficiently visualise and interact. For example, the early stages involve the initial exploration of the data, the middle stages require the validation of methods, and the final stages require the verification and presentation of final results.   

Current solutions for this are often handled through various compromises to make multidimensional data compatible with traditional forms of digital media.  With the introduction and increased accessibility of immersive technologies, such as the digital planetarium \cite{marchetti_adass, sivitilli2023planetarium} and virtual reality \cite{jarrett2021exploring}, an effective alternative means of interfacing with such digital data has become readily available. In the context of big data, this means revealing the particular aspects of the data that traditional media formats hinder during analysis. For example, viewing or selecting areas along the line of sight becomes a challenge with dense data. 

With the added benefits of immersive technology, its novelty presents a lingering challenge. Conventional tools and workflows tend to be confined to the keyboard and mouse. A major exercise is thus avoiding temptations to implement an immersive analogue for every component of the data visualisation and analysis process. This runs the risk of disrupting the workflow of researchers through overloading working memory resources \cite{BADDELEY197447}. Rather, developers must identify what can be kept in traditional form and introduce the alternative where it is needed. In that way, we hope not to disrupt workflows, but complement them for the goal of enhancing their productivity.

\subsection{Other Tools}
In the context of multidimensional data in astronomy, there are a number of existing software solutions that have been developed to visualise and analyse large data sets \cite{Marchetti2024}. 
\subsubsection{CARTA}
The Cube Analysis and Rendering Tool for Astronomy (CARTA) \cite{carta} is a browser-based desktop application for visualising high-volume data sets. It is designed to handle large data cubes and provides a range of visualisation options through colourmaps and contours to represent 3D data on a 2D screen. CARTA is built on a client-server architecture, allowing for efficient data transfer and processing to enable local visualisation of large remote data sets. It supports various file formats, including FITS and HDF5. CARTA also provides a range of analysis tools, including spectral profiling, moment mapping, and masking.
\subsubsection{SlicerAstro}
SlicerAstro \cite{slicerastro} is a package for 3D Slicer, a free and open-source software platform for medical image informatics, image processing, and three-dimensional visualisation. SlicerAstro provides a range of tools for visualising and analysing astronomical data, including support for displaying 3D FITS files through slice, volume and isosurface rendering. It also includes a range of modules for specific tasks, such as spectral analysis, source inspection and kinematic modelling. Unlike CARTA, SlicerAstro can render 3D data sets in a 3D rendered space, but it is also primarily designed for use with a mouse and keyboard interface, which can be cumbersome for large data sets. For example, to make selections of regions of interest, the user must choose a viewing perspective and then use a mouse to draw a lasso around the region of interest to select voxels from the data set. 

\subsubsection{VisIVO}
The Visualization Interface for the Virtual Observatory (VisIVO) \cite{tudisco2025visivo}, developed by INAF, stands out as a well-established platform particularly suited for large-scale volumetric rendering, the visualisation of 2D astronomical maps, and the inspection of time-dependent simulations. It offers both desktop and web-based applications and is integrated with HPC systems and data processing workflows. However, VisIVO is designed for traditional 2D or static 3D screen-based interfaces and does not natively support immersive technologies such as virtual reality.

\subsubsection{FRELLED}
A VR-capable tool, the FITS Realtime Explorer of Low Latency in Every Dimension (FRELLED) \cite{taylor2025frelled} allows volumetric rendering of astronomical data in the Blender 3D computer graphics software. In addition to the desktop interface, there is a VR mode available for the user. As an add-on to the existing Blender interface through Python scripts, the software is primarily designed as a desktop-based 3D visualisation tool. Its rendering  system adapts the 3D data to Blender's own rendering engine as a stack of 2D slices.

\subsection{Motivation and Overview}
A common issue with most of these packages in the context of 3D datasets is their limitation to an interaction interface that is primarily 2D. This means that the user is limited to a flat screen and a mouse/keyboard interface, which can be cumbersome and inefficient for exploring large data sets, particularly over multiple dimensions. This may lead to a cognitive overload for the user, as more working memory resources are dedicated to navigating the data and switching perspectives over multiple views or times. Constantly switching between different views and tools to analyse the data can hamper the user's ability to focus on the scientific goals of the analysis.   

With that underlying our motivations, one promising solution is our immersive Data Visualisation Interactive Explorer (iDaVIE), a Unity-based software suite built from the ground up \cite{sivitilli_adass} for VR that we have officially released as version 1.0. What started as a prototype for visualising 3D data sets at room-scale, giving a user free-roam to walk around and through their data, more interaction and analysis features have been added over the past few years~\citep{Sivitilli2023iDaVIE}. This includes tools of selection, masking, cropping, and various real-time quantitative calculations. Such tools are key to establishing iDaVIE as a useful scientific tool, particularly with the HI radio astronomy community. However, any 3D volumetric data can potentially be used with our software, such as IFU data or even non-astronomical data from other fields.  

In this paper, we present the version 1.0 release of iDaVIE. This includes a look at the scientific context behind its development, an in-depth description of the software itself, the various use cases that have been published, a performance evaluation, as well as ongoing challenges and future work.

\section{Context}

One major outstanding issue in the field of radio astronomy is the interaction with spectral observational data. These usually take the form of 3D cubes with two spatial dimensions and one spectral dimension. Specifically, when observing a field and isolating a specific transition, such as CO or HI, the spectral axis becomes a velocity axis, revealing the kinematics of the gas of interest. Any step along the data reduction process from raw data to final results means losing information that becomes increasingly difficult to retrieve. This can range from the more minor inconvenience of re-downsampling with new factors to going back to the raw data itself, potentially even re-observing. With immersive viewing of multi-dimensional data, one can immediately see the quality of especially kinematically complex data, which in existing 2D viewers is hard if not impossible. 

Considering one of the original main drivers of iDaVIE, the visualisation of spectral line data cubes produced by MeerKAT \cite{meerkat_paper}, in particular observations of atomic neutral hydrogen (HI), there are a variety of challenges that iDaVIE seeks to overcome.  Twenty years ago, data cubes would be limited in size to hundreds of spatial pixels and spectral channels, resulting in cubes on the order of 100 MB because of the limited bandwidth of radio telescope backends \cite[e.g.][]{pisano2002, whisp_paper}.  For such small cubes, it was possible to quickly step through either the spectral channels or spatial slices and scan them by eye to search for sources or to mask regions of interest for further study.  Additionally, one could easily output ``channel maps'' showing emission in multiple channels all in the same figure.  As the backend capabilities of both single-dish and radio interferometers improved, the data volumes and complexity also increased.  In the past 10--15 years, large, single-dish, HI surveys, such as the Galactic All-Sky Survey~\citep{GASS} or
HI4PI~\citep{hi4pi} yielded cubes of $\sim 20\text{--}30$ GB with $\gtrsim 10^6$ spatial pixels and $\sim1000$ spectral channels. This complexity makes it harder to step through individual channels or spatial slices and interact with the data, as well as increasing the computational challenges in viewing such cubes.  Now, with MeerKAT, ASKAP \cite{askap_paper}, the Jansky Very Large Array (VLA) \cite{vla_paper}, and APERTIF \cite{apertif_paper}, the available bandwidths and spectral resolutions are such that astronomers can produce data cubes with the same number of spatial pixels, but up to 32,000 spectral channels \cite{CHILES, LADUMA, wallaby_paper}!  This makes it nearly impossible to assess data quality, search data cubes, and interact with the data while only looking at a single slice of the data at a time.  Today, astronomers need the ability to quickly search large data cubes for sources, draw or adjust masks of these sources, compare existing catalogues of sources with a data cube, in summary: {\it efficiently interact} with the data, and produce images or videos highlighting the scientific results from these cubes for publication.  This is the critical need that iDaVIE addresses.

Traditional tools typically adapt the data to fit their interface constraints rather than designing interfaces around the data's inherent structure. This approach necessitates various compromises. For example, viewing 3D data cubes channel-by-channel (as in CARTA) transforms spatial navigation into temporal progression as the user scrolls through channels sequentially on a flat screen using keyboard and mouse. While practical for 2D displays, this method makes it difficult to perceive complex three-dimensional structures, particularly the kinematic features revealed by spectral data. Other approaches employ depth cues such as volume rendering, motion parallax, scene lighting, and isosurfaces, or enabling additional navigation modes (translation, zoom, pitch, roll) through keyboard modifiers. Selection tools like cloud lassoing (in SlicerAstro) and particle field rendering (in TOPCAT \cite{topcat} or Partiview \cite{partiview}) also rely on such techniques. However, these methods still require users to either mentally reconstruct the 3D structure from multiple 2D views or master complex navigation controls, both of which may increase cognitive load for the interaction  \cite{sweller_clt}, consuming working memory resources \cite{baddeley2000episodic} that would otherwise be available for scientific analysis, pattern recognition, and collaboration.

Where a solution fits into the workflow is key. We identified both a middle area, between the raw data and the analysis, as well as the final stage when capturing imagery for publication. For the middle area, it was important to consider compatibility with data in appropriate formats, as well as the ability to include catalogues from external sources. In addition, it was necessary to be able to export results back into standard formats, so the data can be used further down the pipeline and in subsequent analysis steps. This meant including the ability to import and manipulate masks of the data in addition to points and regions of interest.

In the case of the final stages of the workflow, presentation was important. The ability to generate high-quality imagery or video for publication or outreach, while preserving the scientific integrity of the data, was a significant consideration. Where possible, we have looked to base iDaVIE on existing standards and practices in the astronomy community. This includes the use of FITS \cite{wells1981fits} files, which are a standard format for storing astronomical data, and the use of WCS (World Coordinate System) transformations, which are used to map pixel coordinates in images to celestial coordinates \cite{wcs_fits_paper}. We have also looked at implementing existing standards for data visualisation, such as the use of standard colour maps and rendering techniques. That being said, we also attempted, where possible, to introduce generalised solutions to problems we were looking to solve. The philosophy behind this was to allow other fields to take advantage of the software, casting the net wider for potential users and contributors.

\section{Software}
iDaVIE is built on the Unity \cite{unity} game engine using the SteamVR \cite{steamvr} ecosystem of VR interaction tools. Building the software also requires libraries for handling data from the FITS standard. Specific dependencies for the release version include:
\begin{itemize}
    \item \textbf{Unity} (version 2021.3): Game engine for rendering and interaction.
    \item \textbf{SteamVR} (version 2.7.3): VR device and controller support plug-in for Unity.
    \item \textbf{CFITSIO}  (version 3.410) \cite{cfitsio_paper} and \textbf{AstLib} (version 9.2.14) \cite{ast_paper}: FITS file and coordinate handling in native plug-ins.
\end{itemize}

All dependencies, except for Unity itself, are open-source or freely available for academic use. Readers can refer to the GitHub repository for a complete list of these along with installation instructions (see Section~\ref{label_availability}). Note that these dependencies are built into the Unity project as a standalone executable and are not separately required by the user unless they wish to build the project from source.

A single scene is implemented with Unity in which the user can load and interact with the data set as a VR user while wearing a headset and holding tracked controllers. Behaviour of scene objects, such as the datacube and menus, are managed by C\# scripts attached to these objects. More performance-intensive reading, writing, and manipulation of data are handled using Unity's native plug-in system with imported libraries compiled from C++. As PC-tethered virtual reality (PCVR) software, iDaVIE is designed to be run on a desktop computer with a dedicated graphics card. The software is currently only available for the Windows operating systems as support for graphics drivers, Unity, and PCVR is most robust on this platform. Although a wired connection is found to be the most reliable, iDaVIE is also compatible with wireless setups which allow the user to roam freely in the virtual environment.

\begin{figure}
\centering
\includegraphics[width=\linewidth]{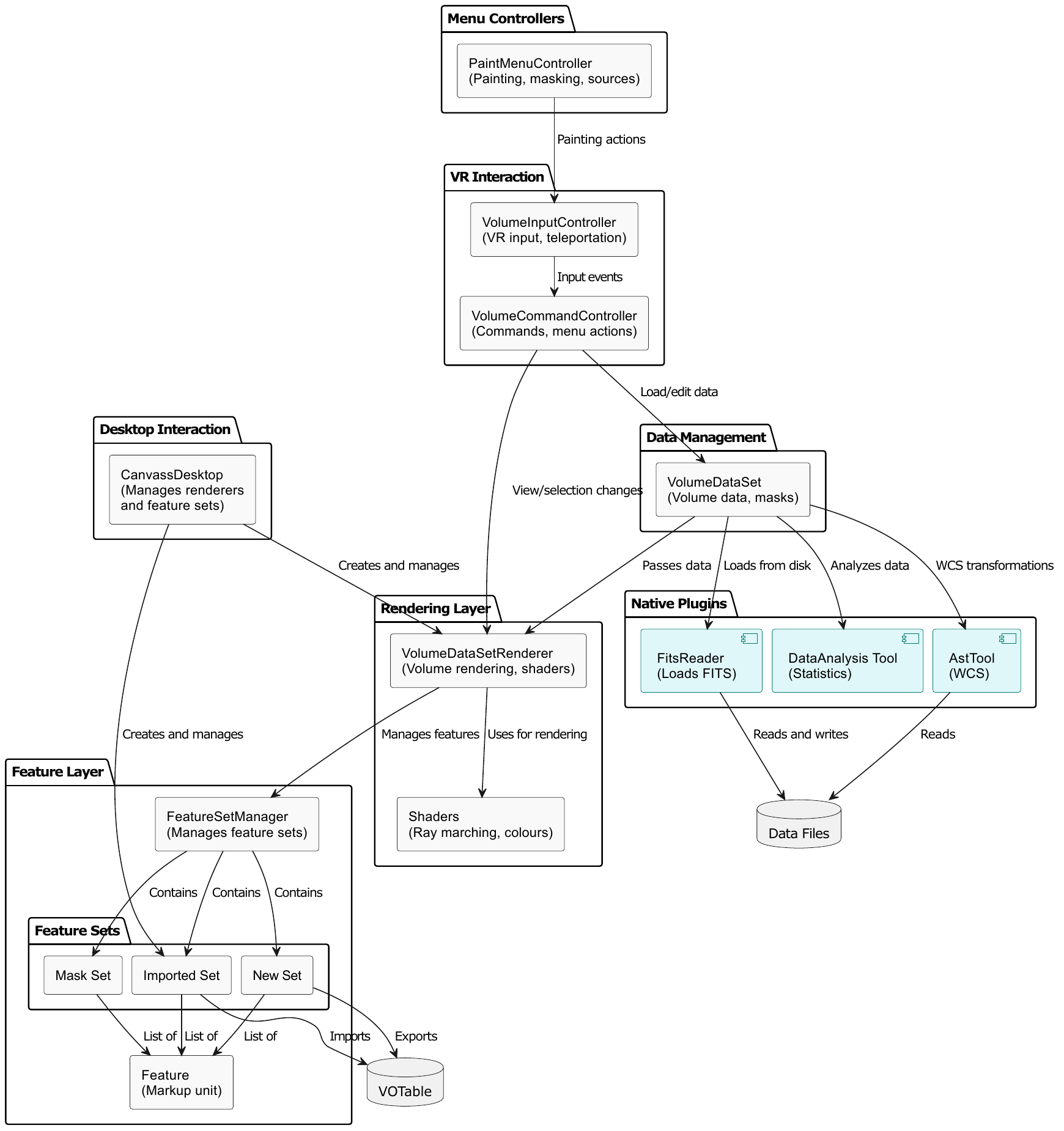}
\caption{This is a simplified high-level architecture of iDaVIE. The various classes and their interactions are shown. Classes are also grouped into different groups based on what component of the software they hold responsibility for.}
\label{fig:architecture}
\end{figure}

Using the \texttt{MonoBehaviour} system of Unity, the single scene contains all the necessary components for rendering and interacting with the data. This scene is populated with several GameObjects, each of which is responsible for one or more specific aspects of the software. This includes the desktop GUI, the interaction system, the feature system, and the rendering layer. The rendering engine is responsible for loading and displaying the data set, while the GUI provides the user with options for selecting and manipulating the data. The interaction system allows the user to interact with the data using their VR controllers. The Feature system allows the user to create and manipulate lists of Features, which are used to mark regions of interest in the data set (see Section~\ref{feature_system}). A simplified high-level architecture of iDaVIE is shown in Figure~\ref{fig:architecture}.

\subsection{Data Handling}
\label{parsing_data}

\subsubsection{I/O Capabilities}

iDaVIE accepts 3D FITS data cubes and optional 3D FITS mask cubes of matching dimensions. FITS files with fewer than three dimensions are not supported, as the software is designed for volumetric data visualisation. When a 4D FITS cube is provided, iDaVIE automatically identifies the spectral (depth) axis if only one of the third or fourth axes has non-unit length. If both axes have non-unit length, the user is prompted to select which axis should be treated as spectral, and the corresponding 3D subcube is loaded. For FITS files with more than four dimensions, any axes beyond the fourth are ignored.

WCS headers are parsed (when present) to map voxel coordinates to sky and spectral coordinates for cursor readouts, selection measurements, and catalogue overlays. The software can export user-edited masks back to FITS, preserving the original cube dimensions and WCS metadata where available, and can export user-defined Features to VOTable catalogues. Cropped subcubes can be saved as FITS files (data and/or mask), with headers updated to reflect the new bounds and retained WCS metadata. These exports are designed for reintegration into downstream analysis pipelines, with the primary constraint that masks must align with the associated data cube dimensions.

\begin{figure}[htbp]
  \centering
  \begin{subfigure}[b]{0.52\textwidth}
    \includegraphics[width=\textwidth]{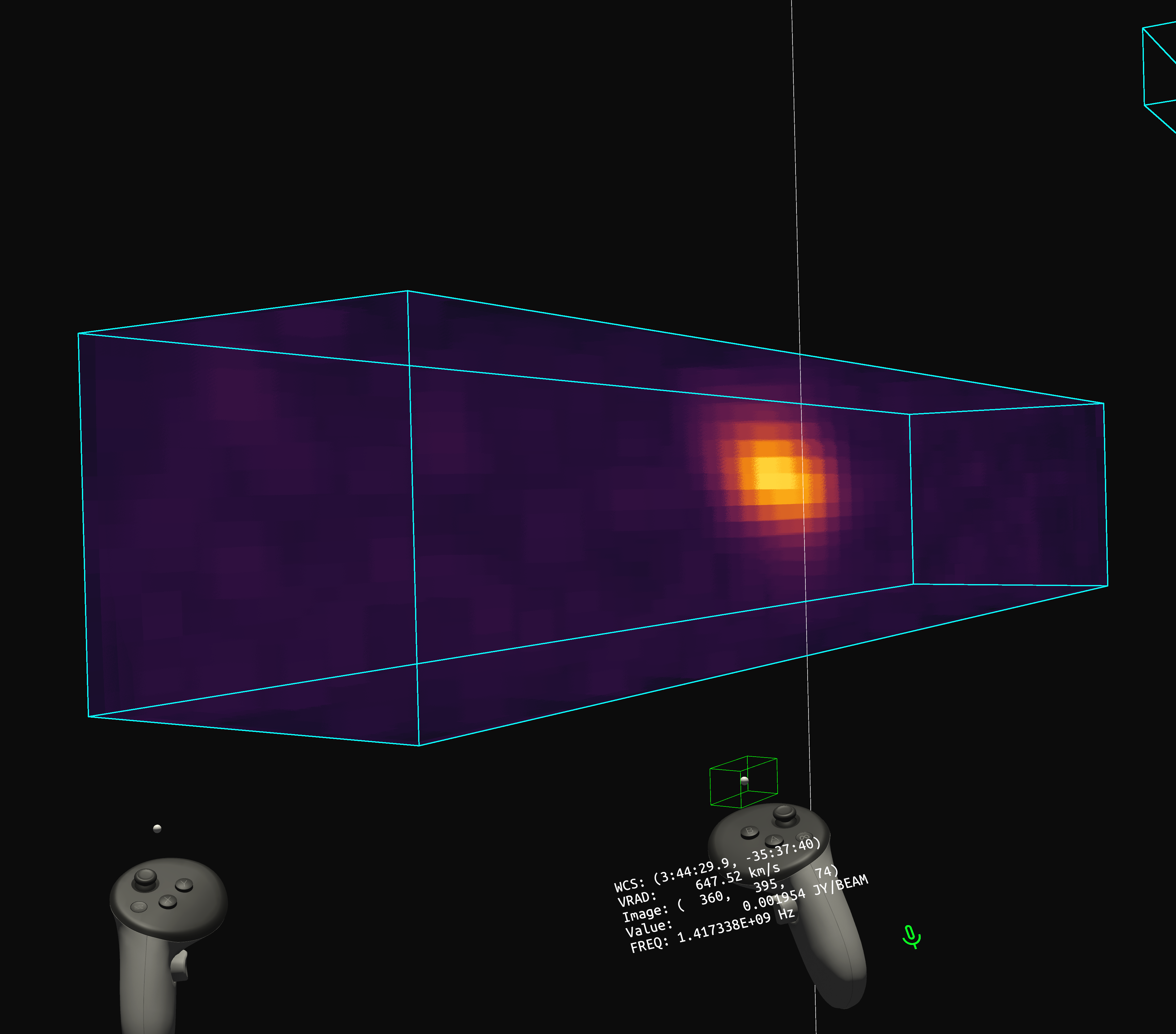}
    \caption{}
  \end{subfigure}\hfill
  \begin{subfigure}[b]{0.44\textwidth}
    \includegraphics[width=\textwidth]{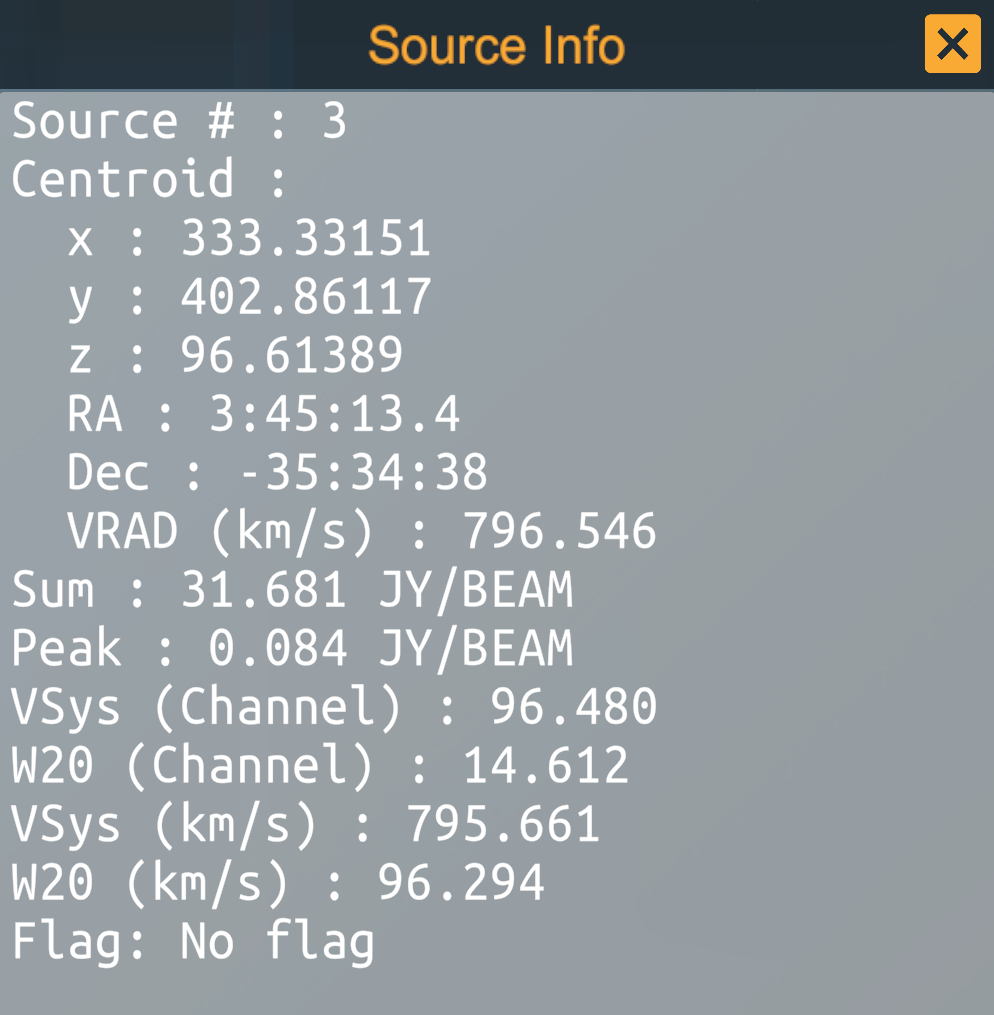}
    \caption{}
  \end{subfigure}

  \vspace{0.6em}

  \begin{subfigure}[b]{0.48\textwidth}
    \includegraphics[width=\textwidth]{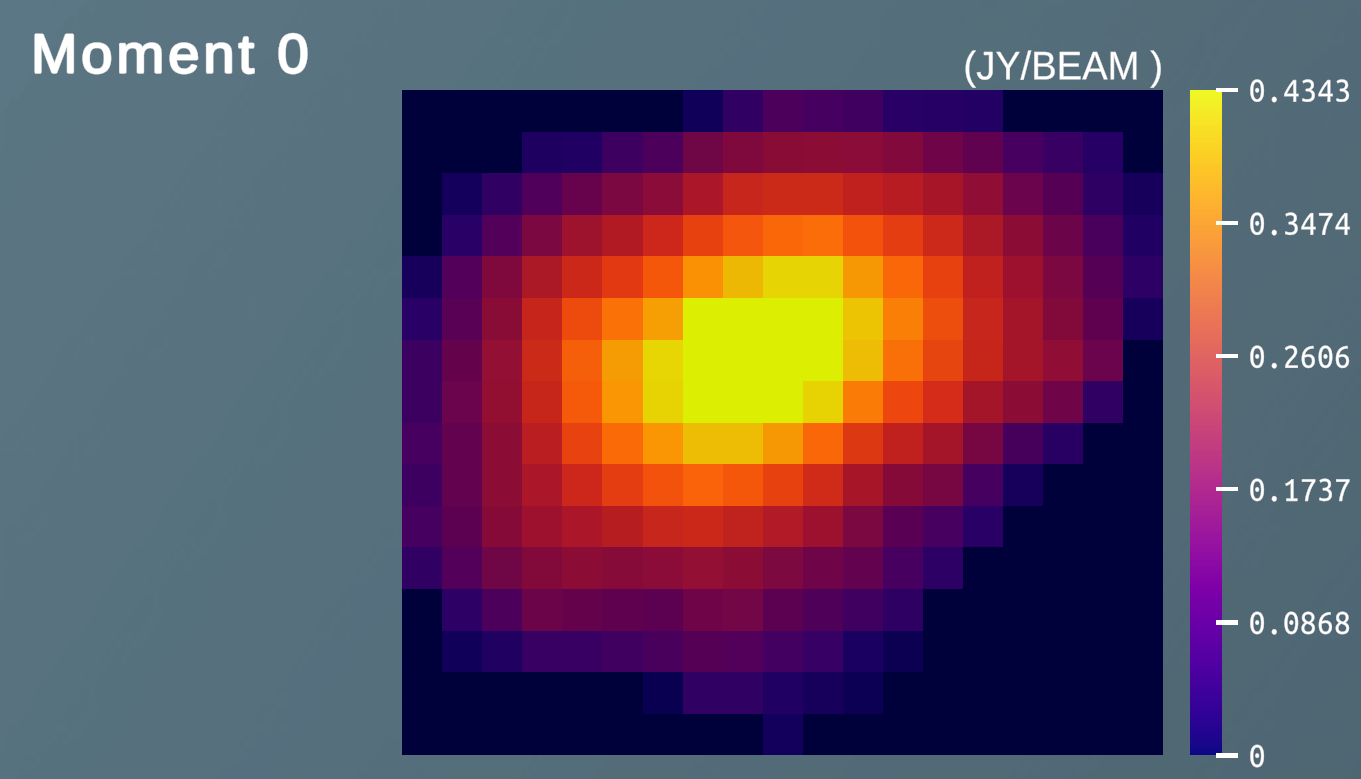}
    \caption{}
  \end{subfigure}\hfill
  \begin{subfigure}[b]{0.48\textwidth}
    \includegraphics[width=\textwidth]{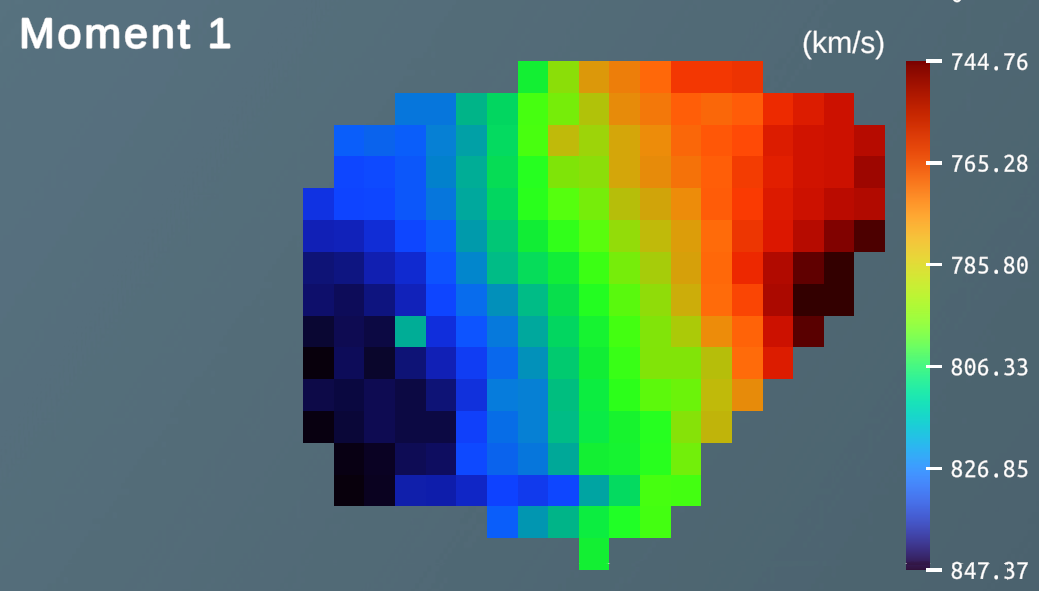}
    \caption{}
  \end{subfigure}

  \vspace{0.6em}

  \begin{subfigure}[b]{0.48\textwidth}
    \includegraphics[width=\textwidth]{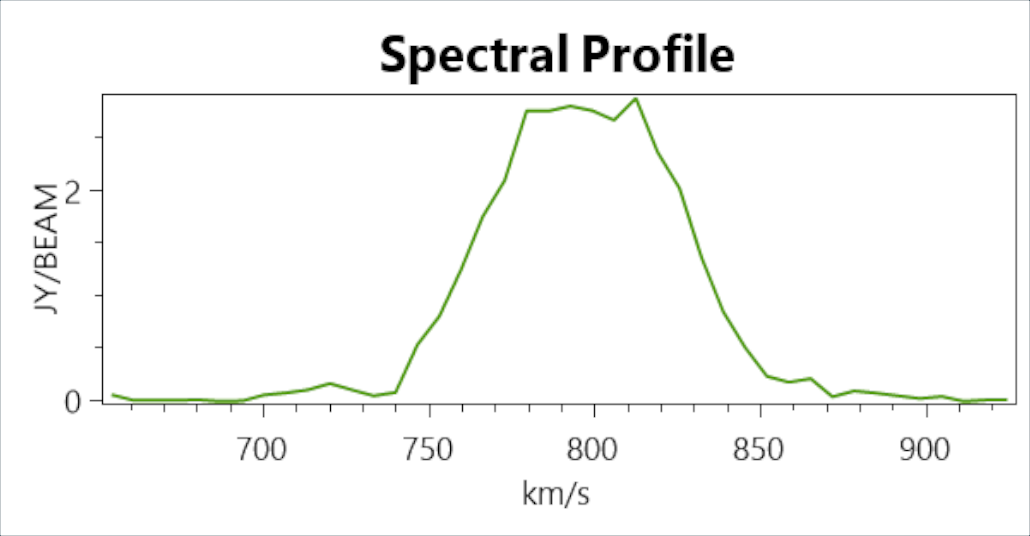}
    \caption{}
  \end{subfigure}\hfill
  \begin{subfigure}[b]{0.48\textwidth}
    \includegraphics[width=\textwidth]{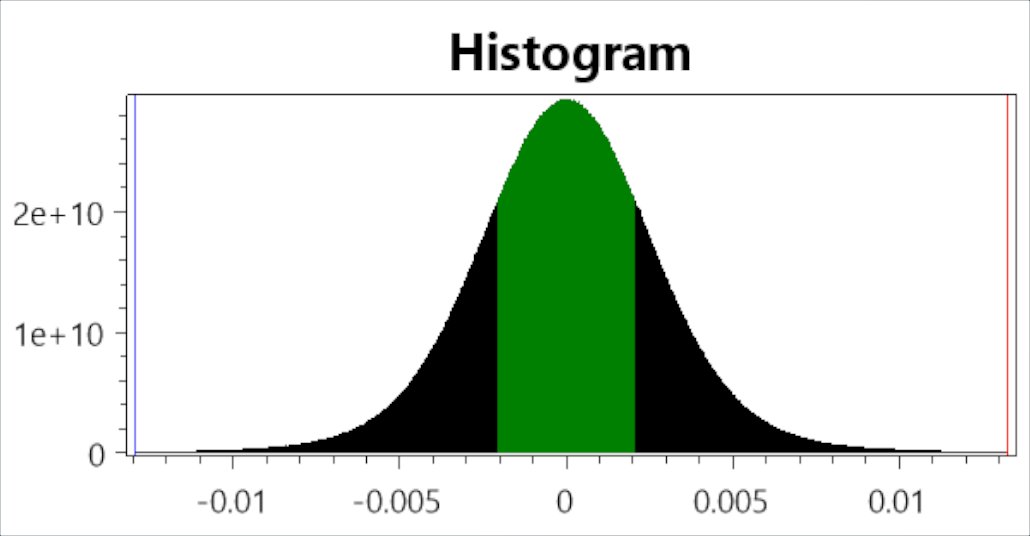}
    \caption{}
  \end{subfigure}

  \caption{Example visualisations generated by iDaVIE in VR from the same data cube and mask: (a) the user’s 3D VR view (exportable as \texttt{.png} via voice command or the \texttt{QuickMenu}); (b) the floating Source Lists info window showing mask-derived statistics or imported VOTable columns; (c) and (d) realtime Moment 0 and Moment 1 maps for cropped regions (exportable as \texttt{.png} or \texttt{.fits}); (e) a spectral profile extracted from selected mask regions (exportable as \texttt{.csv}); and (f) a histogram of the dataset that updates with colourmap clamping.}

  \label{fig:visualisations}
\end{figure}

Furthermore, generated moment maps can be exported as \texttt{.png} images for quick visualisation, and as FITS files with appropriate WCS headers for further analysis. Spectral profiles extracted from selected mask regions can be exported as \texttt{.csv} files containing spectral coordinate (e.g., velocity or frequency) and flux values, with units taken from the cube's WCS header when available. The software also supports exporting high-quality \texttt{.png} images of the rendered data cube for publication or outreach purposes. We present examples of these generated visualisations from the same cube and mask in Figure~\ref{fig:visualisations}.

\subsubsection{Native Plug-ins and Load Pipeline}

In order to efficiently deal with large astronomical data sets and their file standards, iDaVIE uses a set of custom C++ plug-ins that are compiled into a native \texttt{.dll} file. This plug-in is responsible for handling the data parsing and manipulation, as well as providing access to the necessary libraries for reading FITS files, performing World Coordinate System (WCS) transformations, and performing calculations for statistics and downsampling. The plug-ins are designed to be easily integrated into the Unity environment with interface classes written in C\# and accessible by objects within the scene. This allows for seamless communication between the C++ code and the Unity engine. These plug-ins are referred to as \textit{native} plug-ins.

\begin{figure}
\centering
\includegraphics[width=\linewidth]{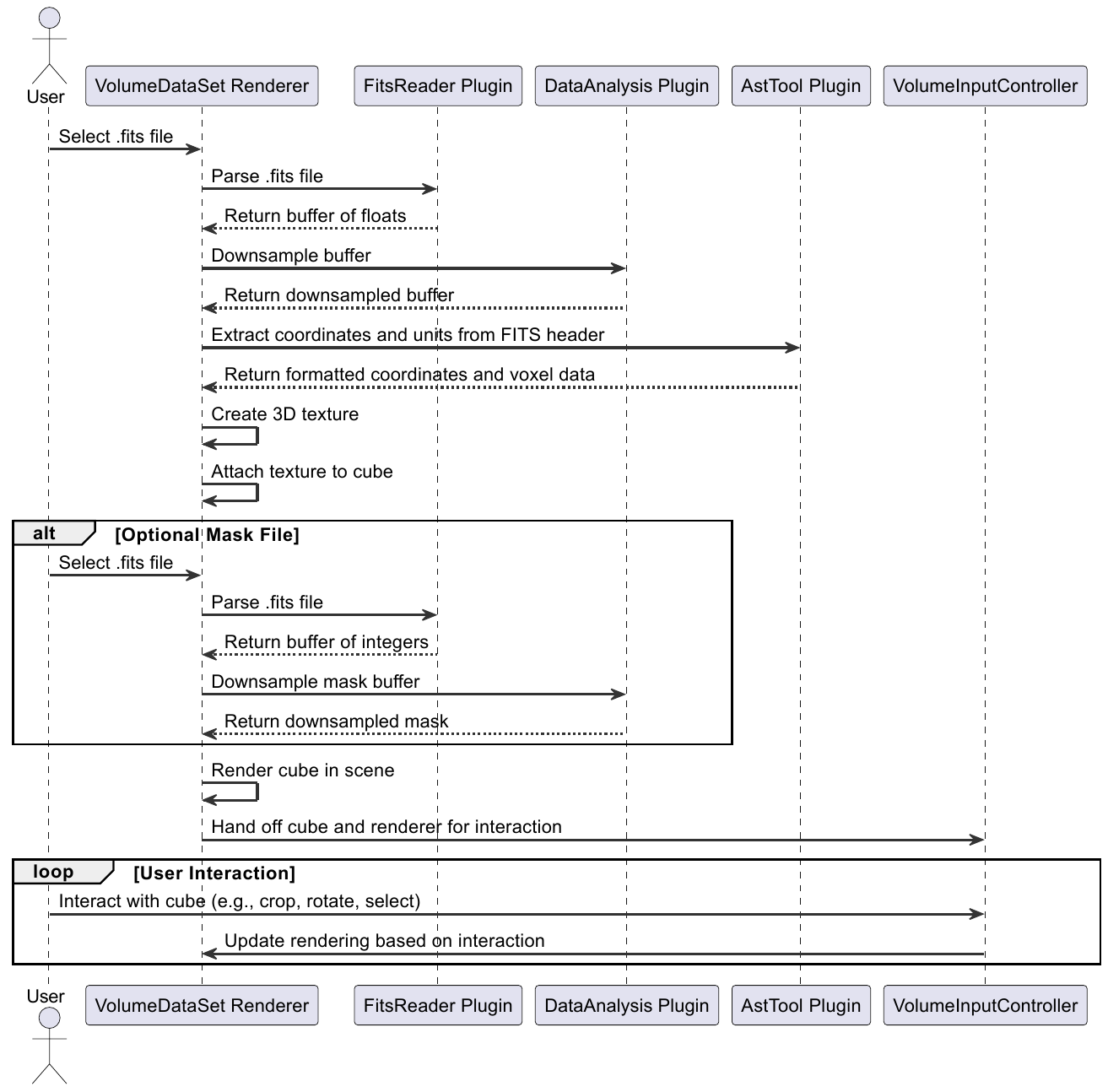} 
\caption{Sequence diagram of the file load process. The user selects a FITS file to load, which is then parsed by the \texttt{FitsReader} plug-in. The data is downsampled and loaded into a 3D texture, which is then rendered in the scene.}
\label{fig:file_load} 
\end{figure}

Upon selecting a file to load, the desktop GUI object spawns a cube object into the scene that represents the data set. The script attached to this cube uses a custom \texttt{FitsReader} native plug-in to parse the selected FITS file into a data object. Upon completion of importing the data from the file, The 3D texture for this cube is then created using the renderer from a buffer of floats supplied by the plug-in and downsampling via the method described below. The user can optionally select a mask file to be loaded into the scene. This is a 3D FITS cube of the same dimensions as the data cube, but with short integer values to indicate the presence of a masked region. The mask file is loaded into the renderer as its own data object and also converted to a 3D texture through the same downsampling method. With the data cube and mask cube loaded as 3D textures, a material with an attached volume-rendering shader (see Section~\ref{volumetric_rendering}) is loaded. The result is a cube in the scene with an appearance of being composed of voxels whose colour is representative of the voxel values of the loaded datacube. The cube and its attached renderer are then handed off to a controller object in the scene to handle interaction. A sequence diagram of the file load and data parsing process is shown in Figure~\ref{fig:file_load}.

The 3D texture size must be limited. The first reason is due to a 4GB hard-limit set by Unity with 32-bit memory addressing. The second reason comes from rendering bottlenecks. The default limit, which we optimized through trial-and-error for user comfort, is conservatively set to 368 MB, but this can be adjusted in a configuration file for higher-power machines with more powerful GPUs. In order to assure a 3D texture size smaller than this limit, the buffer undergoes downsampling from our custom \texttt{DataAnalysis} native plug-in. This is accomplished through a parallelised block-based aggregation method. Three downsampling factors are supplied to the plug-in for each axis of the data cube. The plug-in then calculates the new dimensions of the data cube and creates a new buffer with the downsampled data. The downsampling factors are determined by starting at 1,1,1 and increasing them until the new dimensions achieve a size smaller than the texture limit. The $x$ and $y$ axes factors are locked as they are usually the spatial axes. The increase happens through increments of 1, first for $x$-$y$, then $z$, then $x$-$y$ again, and so on. The voxels of the new cube then become either the average or maximum value of the voxels in the analogue 3D window in the original cube, depending on the selected downsampling method indicated in the configuration file. If the user crops the data to a 3D volume that is smaller than the texture limit (necessary for voxel-based interaction with the data) the downsampling step is skipped. The full data cube and mask are retained in RAM for when the user crops to other regions of interest. This allows users to enhance subregions in VR and switch back to the lower‑resolution full cube or other regions as required, without reloading data from the file.

Our custom \texttt{AstTool} plug-in makes use of the AstLib library to extract coordinates, units, and other important information from the header of the FITS file. This is also done during the load process by taking the header of the FITS file and setting the formatting and units of both the coordinates and voxel data values. 

\subsection{User Interaction}

The Graphical User Interface (GUI) is built using Unity's native UI system. The GUI is designed to be intuitive and user-friendly, allowing easy navigation through the various features and functionalities of the software. In planning the GUI, we have taken into consideration the needs and preferences of the target users, which include astronomers and data scientists, particularly in the HI community. This also includes various decisions regarding what features to expose on the desktop (Desktop GUI), what to expose in the VR environment, particularly in what mode (controller states, voice commands, VR GUI) and what to share between the two. 

For the VR interaction, we rely on the SteamVR plug-in for Unity as the platform offers flexibility and stability. It makes use of the OpenVR standard, allowing iDaVIE to be compatible with most modern VR headsets, including the HTC Vive, Valve Index, and Meta Quest series. It also provides a controller-binding interface with labelled actions that can be mapped to the buttons of any compatible controller. This allows not only scaling to future or niche headsets, but also easy customisation of the controls to suit the user's preferences. A performance diagnostic tool showing frame timing and GPU/CPU load for the headset also allow for effective debugging and optimisation during development. Additionally, the SteamVR ecosystem continues to be actively developed and maintained by Valve, supporting the long-term viability of iDaVIE.

\subsubsection{Desktop GUI}
\begin{figure}
  \centering
  \includegraphics[width=1.0\textwidth]{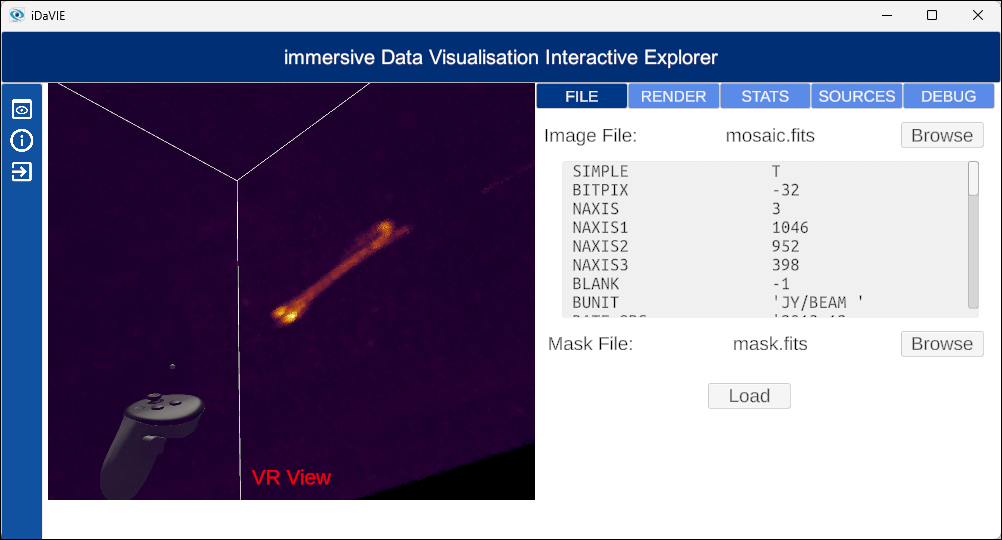}
  \caption{The file tab of the desktop GUI. This is what the user first sees when loading iDaVIE. Browsing for a data or mask file opens the system default file browser and filters for FITS files. A scrollable text box gives a preview of the data file FITS header. The VR View box shows a duplicate of what the VR headset is currently rendering. }
  \label{fig:desktop}
  \end{figure}

The desktop GUI in iDaVIE (see Figure~\ref{fig:desktop}) is designed to provide astronomers with a familiar and efficient interface for managing data and visualisation parameters. Rather than replicating every function in VR, the desktop GUI focuses on tasks that benefit from conventional input methods, such as file selection, parameter tuning, and reviewing debug messages. This approach aims to minimise cognitive load by avoiding the introduction of unnecessary interaction paradigms and maintains compatibility with established workflows \cite{juliano2022increased}.

Key features for the desktop include data and mask file-loading, visualisation parameter adjustment, and real-time rendering previews. The GUI also provides access to configuration options such as colourmap selection and clamping, in addition to rest frequency settings, with changes reflected immediately in the VR environment. Error handling and feedback are integrated to ensure robust operation, while advanced or less frequently used options are accessible but unobtrusive.

By separating desktop and VR functionalities, with intentional overlap where appropriate, iDaVIE leverages the strengths of each environment, allowing users to efficiently prepare and adjust their data before immersive exploration. For a comprehensive guide to all GUI options and workflows, users are referred to the online documentation (see Section~\ref{label_availability}).

\subsubsection{Controller States}

\begin{figure}
  \centering
  \includegraphics[width=0.9\linewidth]{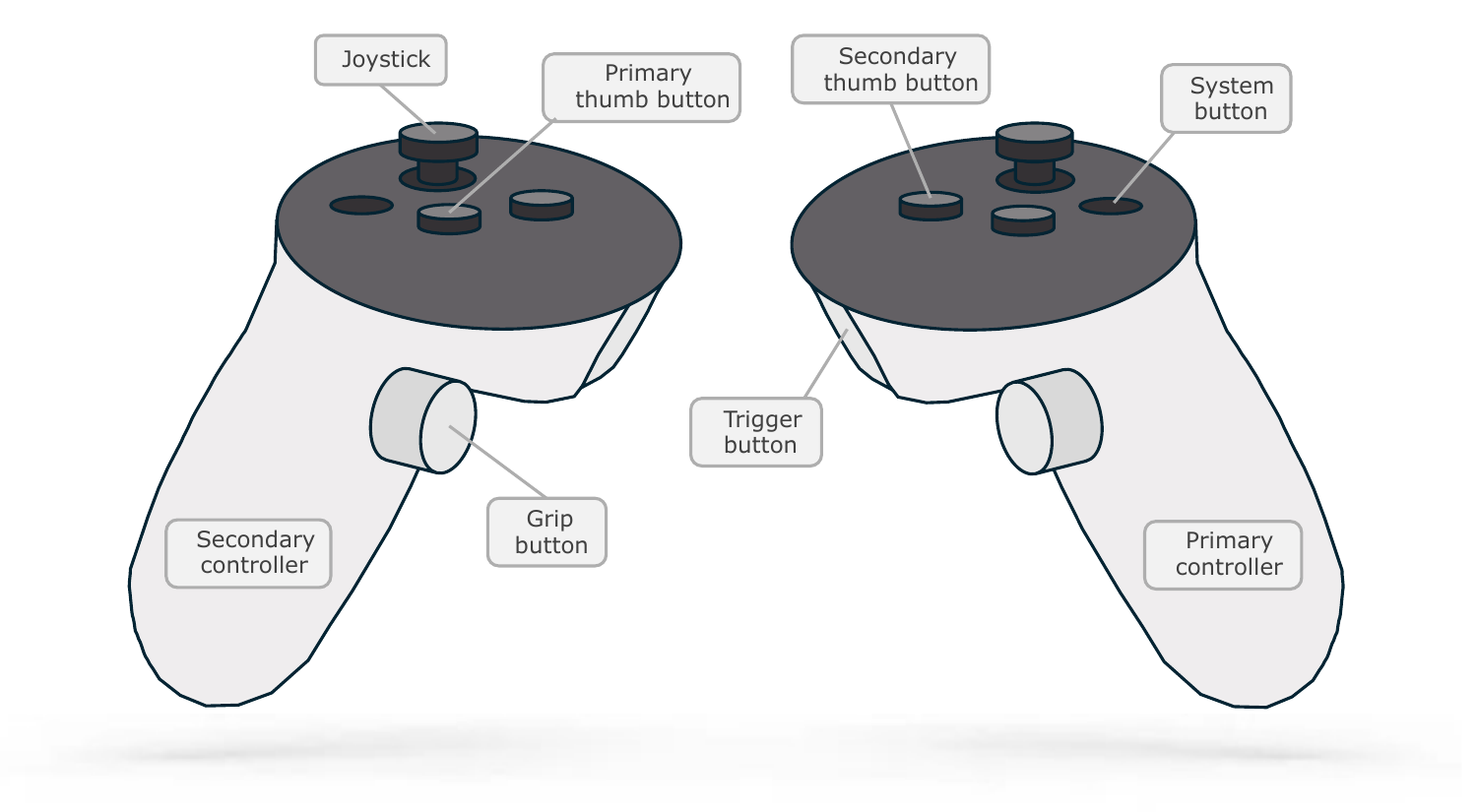}
  \caption{Layout of the input buttons referenced in this paper for the Meta Quest 3 controllers. This layout is mapped to other VR controllers via the SteamVR action system, allowing custom button bindings across devices. Buttons are labelled once here but apply to both controllers. The primary controller is the user's dominant hand, set in the VR Settings menu.}
  \label{fig:controller_layout}
\end{figure}

The user's headset and controller position, along with the various button inputs and haptic outputs, are managed by a SteamVR player object in the scene. A custom script, attached to the same player object, manages how the user as a VR-player interacts with the scene, specifically with the spawned volume cube. This is implemented with two state machines, the \texttt{LocomotionState} machine for navigating around the cube and the \texttt{InteractionState} machine for interacting with the cube. A diagram of the VR controller layout with button labels is provided in Figure~\ref{fig:controller_layout} for reference.

\begin{figure}
  \centering
  \begin{subfigure}[b]{0.52\textwidth}
    \includegraphics[width=\textwidth]{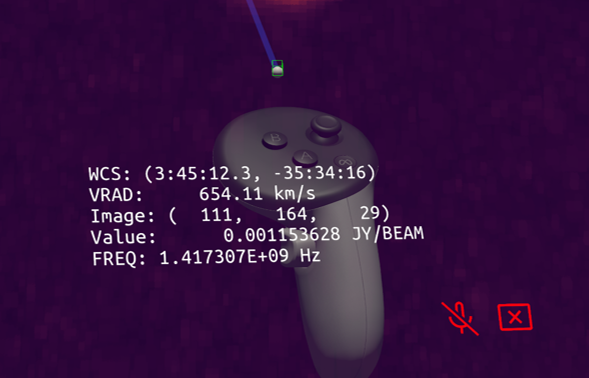}
    \caption{}
  \end{subfigure}\hfill
  \begin{subfigure}[b]{0.44\textwidth}
    \includegraphics[width=\textwidth]{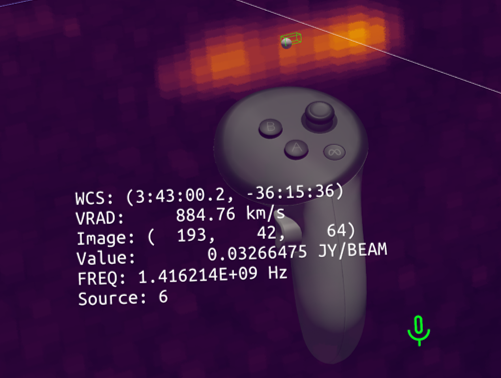}
    \caption{}
  \end{subfigure}

  \vspace{0.6em}

  \begin{subfigure}[b]{0.52\textwidth}
    \includegraphics[width=\textwidth]{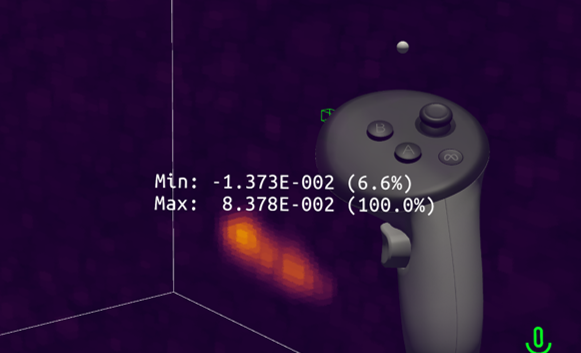}
    \caption{}
  \end{subfigure}\hfill
  \begin{subfigure}[b]{0.44\textwidth}
    \includegraphics[width=\textwidth]{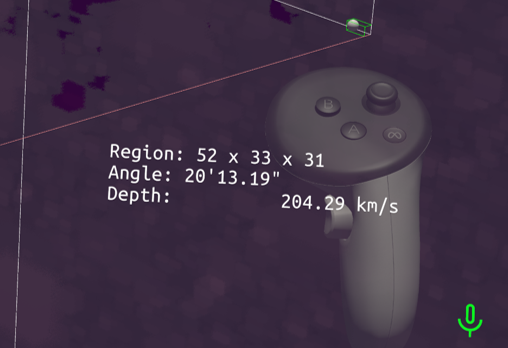}
    \caption{}
  \end{subfigure}

  \caption{The \texttt{CursorInfo} object displays: (a) and (b) the controller cursor position in world and data coordinates, plus voxel value and mask region number (``Source'') when applicable; (c) parameter-adjustment feedback (e.g., colourmap clamping); and (d) selection-box feedback. In all panels, the microphone icon indicates voice recognition status (green active, red inactive).}
  \label{fig:cursor_info}
\end{figure}

\begin{figure}
  \centering
  \includegraphics[width=0.55\linewidth]{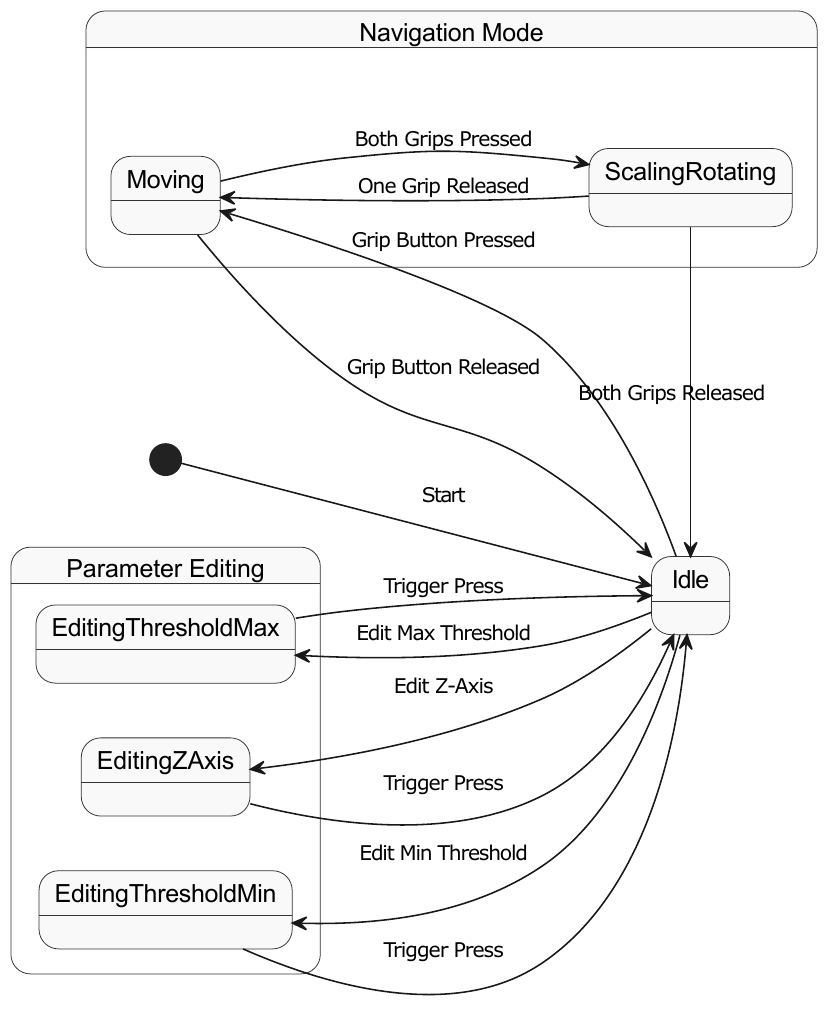}
  \caption{The \texttt{LocomotionState} machine. The user starts in the \texttt{Idle} state, where they can move the controller to update the \texttt{CursorInfo} object. The user can then enter the \texttt{Moving} state by holding the grip button of a single controller, or the \texttt{Scaling}/\texttt{Rotating} state by holding the grip button of both controllers. If entering the \texttt{Parameter Editing} state through a voice command (see Section~\ref{voice_commands}), the user can adjust the value of the parameter by moving the controller up and down. The state is exited (and parameter adjustment saved) by pressing the trigger button.}
  \label{fig:locomotion_state}
  \end{figure}

For the \texttt{LocomotionState} (see Figure~\ref{fig:locomotion_state}), the user starts in the \texttt{Idle} state where the controllers can be moved into positions, providing feedback to the user based on position in space via a text indicator on the user's controller. This displays text attached to the primary controller (see Figure~\ref{fig:cursor_info}). This is updated with various information about the controller's cursor in the scene, including position coordinates, the corresponding transformed coordinates in data space, the value of the voxel the cursor is within, and the integer number of the mask region if the cursor lies within one.

Users can navigate through the data by ``grabbing'' space using the grip buttons on their VR controllers. When a grip is held on one controller, translating the controller in space results in a corresponding translation of the data cube, creating the sensation of physically pulling or pushing the data environment. This interaction acts as a three-dimensional analogue to swiping across a touchscreen. When both grips are held simultaneously, users can rotate or scale the data cube by moving their hands apart or together, offering a natural 3D equivalent to ``pinch-to-zoom'' and rotation gestures familiar from touchscreen interfaces.

In addition to spatial manipulation, users can enter a parameter adjustment mode via voice command. Within this mode, parameters such as colourmap clamping or cube size proportions can be modified by moving the controllers vertically. This direct mapping of hand movement to parameter change enables precise and intuitive fine-tuning of the visualisation settings.

\begin{figure}
  \centering
  \includegraphics[width=\linewidth]{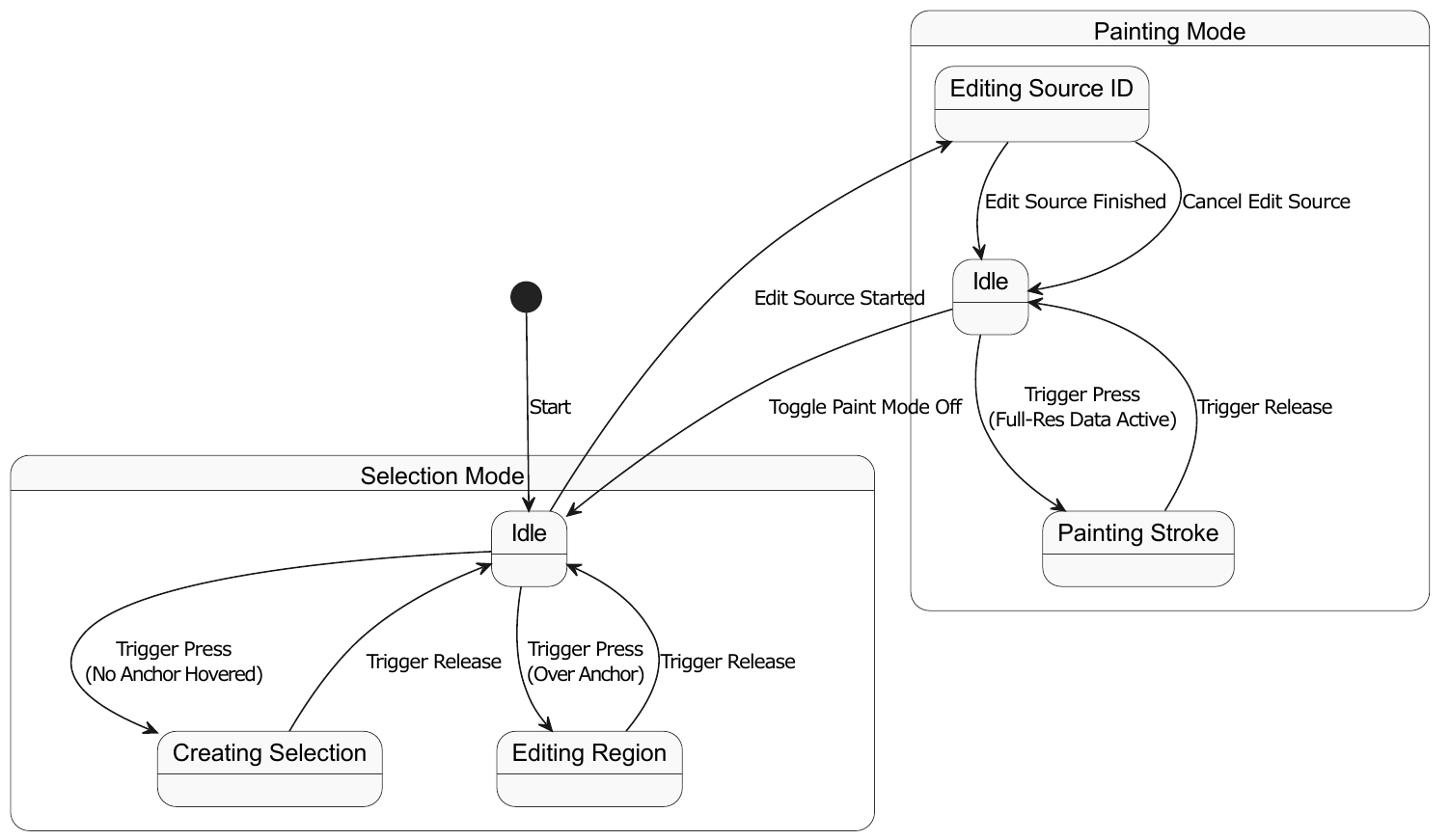}
  \caption{The \texttt{InteractionState} machine. The user starts in the \texttt{Idle} state of Selection Mode, where they can move the controller to see the \texttt{CursorInfo}. The user can then enter the \texttt{Creating} state by holding the primary button of the primary controller in empty space, or \texttt{Editing} state by doing the same in the anchor point of an existing Feature. When entering Paint Mode through the GUI or voice command, the user starts in the \texttt{Editing Source ID} state where they either select a masked source or create a new one. They then enter \texttt{Idle} state of Paint Mode. They enter the \texttt{Painting Stroke} state by holding the primary button of the primary controller.  The state is exited (and painting saved) by exiting Paint Mode in the Paint Menu.}
  \label{fig:interaction_state}
  \end{figure}

For the \texttt{InteractionState} (see Figure~\ref{fig:interaction_state}), the user also starts in the \texttt{Idle} state of a Selection mode where the user can move the controller into position to start a box selection of a region within the cube.

The user can enter two types of Selection state from \texttt{Idle}. The first is if they hold the primary button on the primary controller, they enter \texttt{Creating Selection} state for the selection mode. Here a new selection box is spawned into the scene as a special case of a Feature (see Section~\ref{feature_system}) where the first corner is anchored where the user initially pressed the button. The opposite corner of the box is attached to the cursor and follows it until the user releases the button. The result is a temporary Feature box overlaying voxels in the scene, which indicates a ``selected'' rectoid of voxels. The info on the cursor displays the size of the box in voxel dimensions as well as the angle across the sky and depth in frequency or velocity units if WCS information is included in the FITS file.

\begin{figure}
  \centering
  \includegraphics[width=1.0\textwidth]{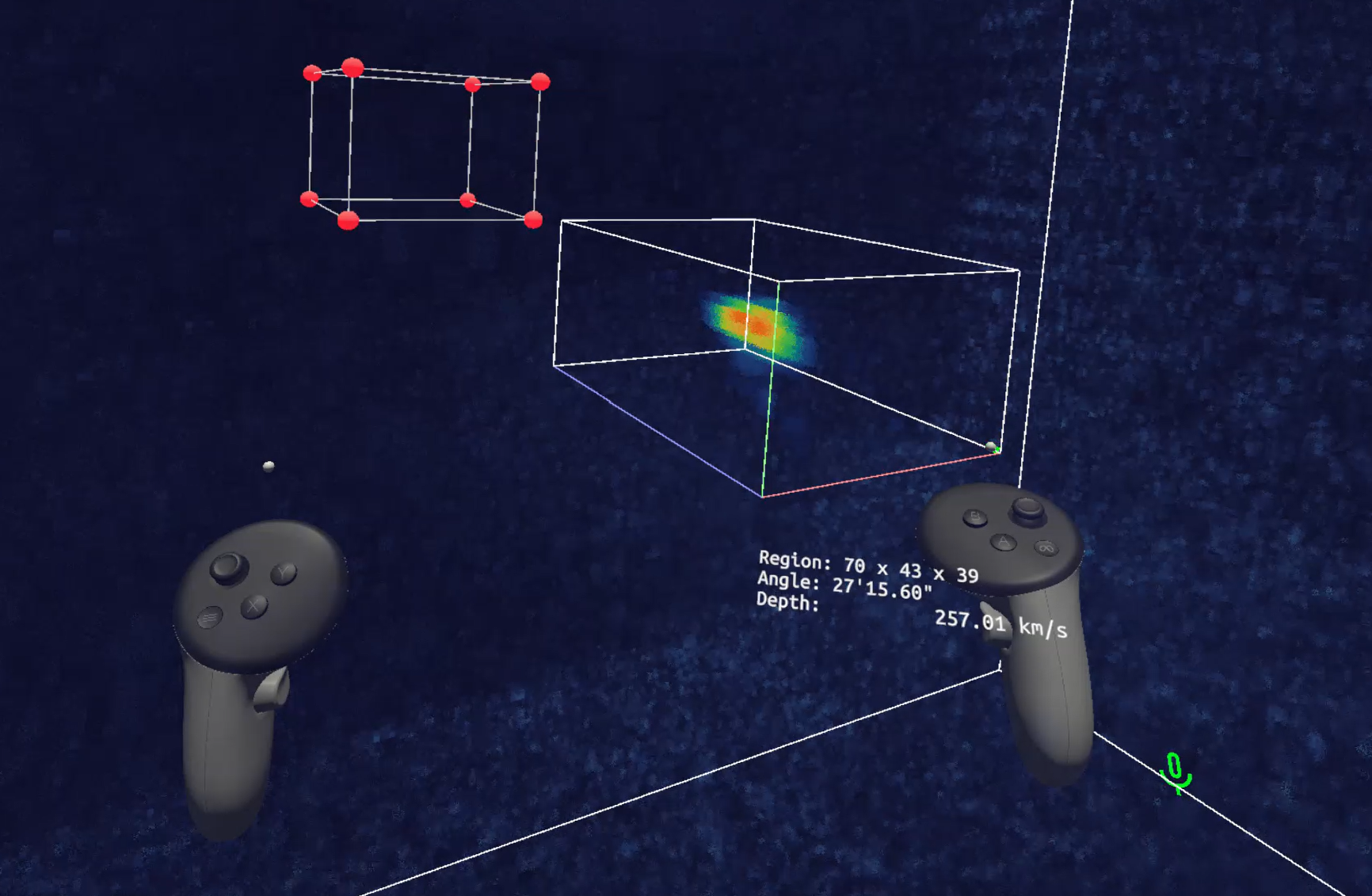}
  \caption{The \texttt{Creating Selection} state of the controller spawns a selection box over which the user can select a group of voxels in the datacube. Information displayed on the controller cursor indicates the dimensions of the selection box, which can also be used as a spatial measuring tool. Colours on the box indicate axis dimensions where $x$ is red, $y$ is green, and $z$ is blue. Note the previous selection box in the background with the anchor spheres available for the \texttt{Editing Region} state. This previous box will disappear upon completion of the new one.}
  \label{fig:selection_tool}
  \end{figure}

The second Selection state type is if the user presses and holds the same primary button of the primary controller, the \texttt{Editing Region} state is triggered, which attaches that corner of the Feature to the cursor, allowing the user to adjust the size of the box. The state is exited upon releasing the button with the Feature's adjusted corner planted in position and the user returns to \texttt{Idle} selection state. The selecting tool is shown in Figure~\ref{fig:selection_tool}. With a selection box present in the scene, the user can crop the rendered data to the selection with either a voice command or using the \texttt{QuickMenu} on the secondary controller (see Figure~\ref{fig:quckmenu} and Section~\ref{vr_gui}).

A key component of iDaVIE is the mask painting interaction, which allows the user to flag individual voxels with an integer. This effectively allows loaded (or new) masks of the data to be edited in realtime by the user. As mentioned previously, the mask is loaded into the scene as a second set of voxel values corresponding to the mask itself. Voxels with integer values of 0 indicate that the voxel is masked out, and non-zero values indicate the voxel is not masked, in which case the value corresponds to whatever the user assigns meaning. We refer to these mask values as ``sources''. This is inspired by the masking system of the HI Source Finding Application (SoFIA) \cite{sofia_paper, sofia_paper_2} source-finding algorithm, which assigns 0 to non-sources and integers to individual galaxy sources. If a mask is not loaded with the data, paint mode creates a new mask that can be saved as a new FITS file.

\begin{figure}
  \centering
  \includegraphics[width=1.0\textwidth]{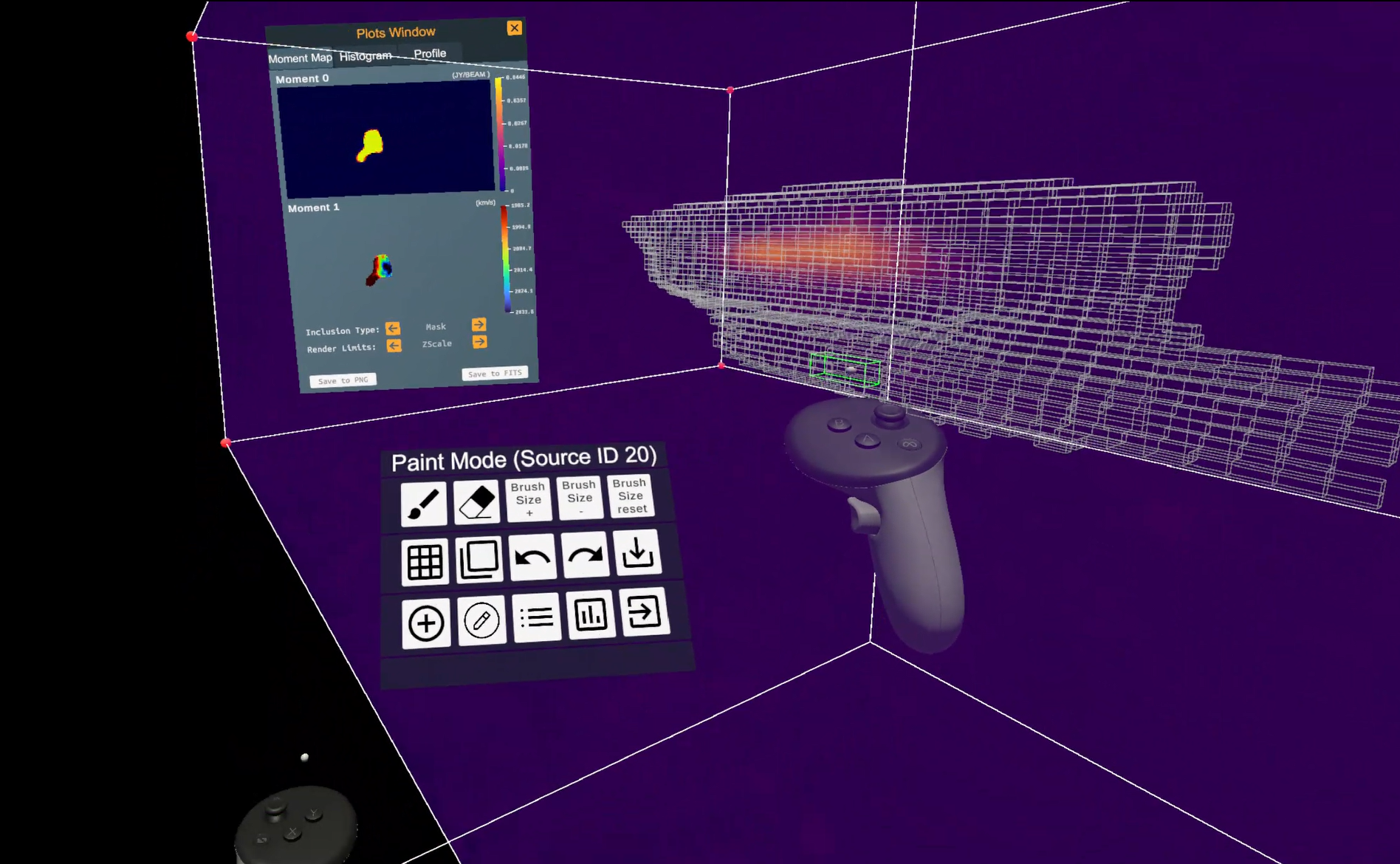}
  \caption{In Painting Mode, a special menu is attached to the user's secondary hand controller and voxels with non-zero mask values are outlined in the scene. The primary hand controller contains a ``paintbrush'' at the cursor that can be adjusted in size with the joystick. When holding the primary thumb button, the user can toggle the mask-status of voxels in the cube. Here this tool is being used alongside the Plots Window for checking how the new mask affects the resultant moment maps.}
  \label{fig:paint_mode}
  \end{figure}

If the user enters Painting Mode via the \texttt{QuickMenu} or a voice command, they start in a \texttt{Source Editing} state where they either move their cursor into a masked region and click the trigger button, or they select ``New Source'' in the Paint Menu to set what Source ID integer they are painting over voxels. This then puts the user in the \texttt{Idle} painting state where they can move the cursor around the cube and adjust the brush size. When the user presses and holds the primary button of the primary controller, the \texttt{Paint Stroke} state is entered where voxels of the mask under the brush of voxels are converted to the value of the set Source ID integer. Upon releasing the button, the user returns to the \texttt{Idle} state of paint mode. If the user exits Paint Mode, they return to \texttt{Idle} of Selection Mode. A rendered wireframe overlays voxels in Paint Mode to indicate which voxels are masked. Voxels with mask values corresponding to the current set Source ID are brighter than other masked voxels. We found this provided the highest contrast and precision over meshes or voxel colours to show the user immediate feedback of their mask painting. The Painting Mode is shown in Figure~\ref{fig:paint_mode}.

\subsubsection{Voice Commands} \label{voice_commands}
\label{sec:voice_commands}
A script is attached to the user object in the scene to provide input for voice commands. This uses the \texttt{KeywordRecognizer}, a Unity wrapper for the speech accessibility tools of Windows. The tool is provided a list of written phrases to listen out for with a confidence sensitivity level set in the configuration file. When a phrase is recognised, a method is triggered. This list includes actions such as cropping the cube to the current box selection or changing the colour map, or triggering a new state, such as adjusting the colour map thresholds with the controller heights. A full list of the voice commands can be displayed to the VR user via a button in the \texttt{QuickMenu}.

To make the voice interaction more robust, we added an indicator to the controller to indicate if the voice recognition system is active or not. This is useful for letting the user know if the iDaVIE desktop window is currently active on the desktop, which is a Windows-level security requirement for the voice recognition system. If the window is active, a green microphone appears to indicate that the voice command system is active and waiting for commands. Otherwise, a red crossed-out microphone appears alongside a window icon to indicate that the iDaVIE desktop window needs to be activated (e.g. see the panels in Figure~\ref{fig:cursor_info}). There is an optional ``push-to-talk'' setting that can be activated in the configuration file that keeps the voice command system inactive with a red crossed-out microphone icon until the user holds in the secondary button of the primary controller. This makes the icon green, indicating that the voice command system is active and ready. This is especially useful in noisy environments where the voice command system might pick up unintentional recognitions in the background. 

In order to give feedback to the user that a voice command was successful, haptic feedback is given in the primary controller via a short vibration. In addition, a ``toast'' pop-up text window is temporarily displayed in the user's field of view with a message to indicate that the voice command was triggered successfully. 

\subsubsection{VR GUI}
\label{vr_gui}
A system of menus and buttons is implemented in the VR environment to allow the user to interact with the scene and data set. This is built on the same Unity UI system as the desktop GUI, but with a few modifications to make it more suitable for VR interaction. The menus are designed to be easily accessible and visible in the VR environment, allowing the user to quickly navigate through the various options and settings. When a ray cast from the controller intersects a menu, a laser pointer is displayed to indicate the intersection. The user can then press the trigger button to select interactable menu items, such as buttons and toggles. Scrollable lists also allow the user to move the controller joystick up and down to quickly scroll through the list of items.

\begin{figure}
  \centering
  \includegraphics[width=1.0\textwidth]{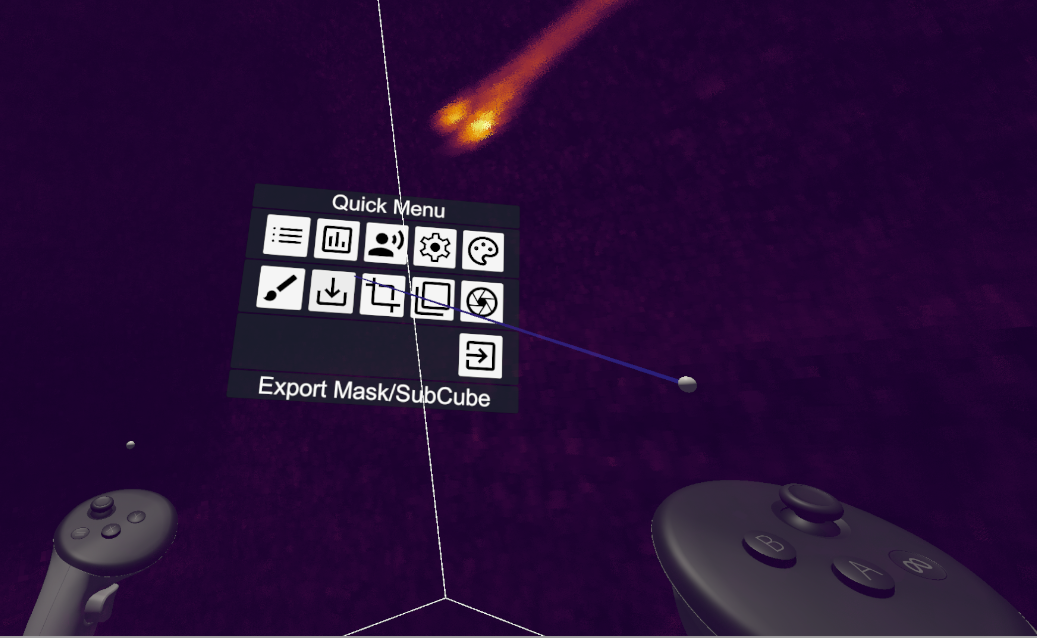}
  \caption{The main method to spawn in floating menus and trigger quick actions in the scene is done with the \texttt{QuickMenu} attached to the user's off-hand controller. Selections can be made with a laser point cast from the primary controller.}
  \label{fig:quckmenu}
  \end{figure}

The main menu of interaction is the \texttt{QuickMenu} (see Figure~\ref{fig:quckmenu}), which is spawned as an attached object on the secondary controller when holding the secondary button. The menu consists of a series of icons that represent different actions and features of the software. The user can select an icon by pointing the controller at it and pressing the trigger button. The menu can be closed by releasing the secondary button.

Multiple floating menus can be spawned in the scene, each with its own attached dedicated Unity \texttt{MonoBehaviour} derived class. These menus can be moved by using the laser pointer to select and hold the top of the menu window and moving the controller to position or moving the joystick to push or pull the window closer to the user. This allows decluttering of the scene and placing menus in convenient locations for the user when necessary

The Feature Menu enables users to manage regions of interest—termed ``Features'' which can be imported from mask files, external catalogues, or created interactively in VR (see Section~\ref{feature_system}). Features are visualised as overlays within the data cube, and can be selected, edited, or grouped for further analysis.

The Plots Menu provides real-time visualisations, including moment maps, histograms, and spectral profiles for selected regions, supporting rapid quantitative assessment without leaving the immersive environment.

Settings and customisation options are accessible through other dedicated menus, allowing users to adjust rendering parameters, colour maps, and interaction modes. Additional menus, such as the Voice Command Window and Paint Menu, further streamline complex tasks like mask editing and command discovery for the user.

This menu-driven approach, combined with VR-native interaction and voice control, allows astronomers to efficiently explore, annotate, and analyse large multidimensional datasets in a manner that complements traditional workflows. For detailed operational instructions, users are referred to the online documentation (see Section~\ref{label_availability}).

\subsection{Volumetric Rendering}
\label{volumetric_rendering}

\begin{figure}
\centering
\includegraphics[width=0.7\textwidth]{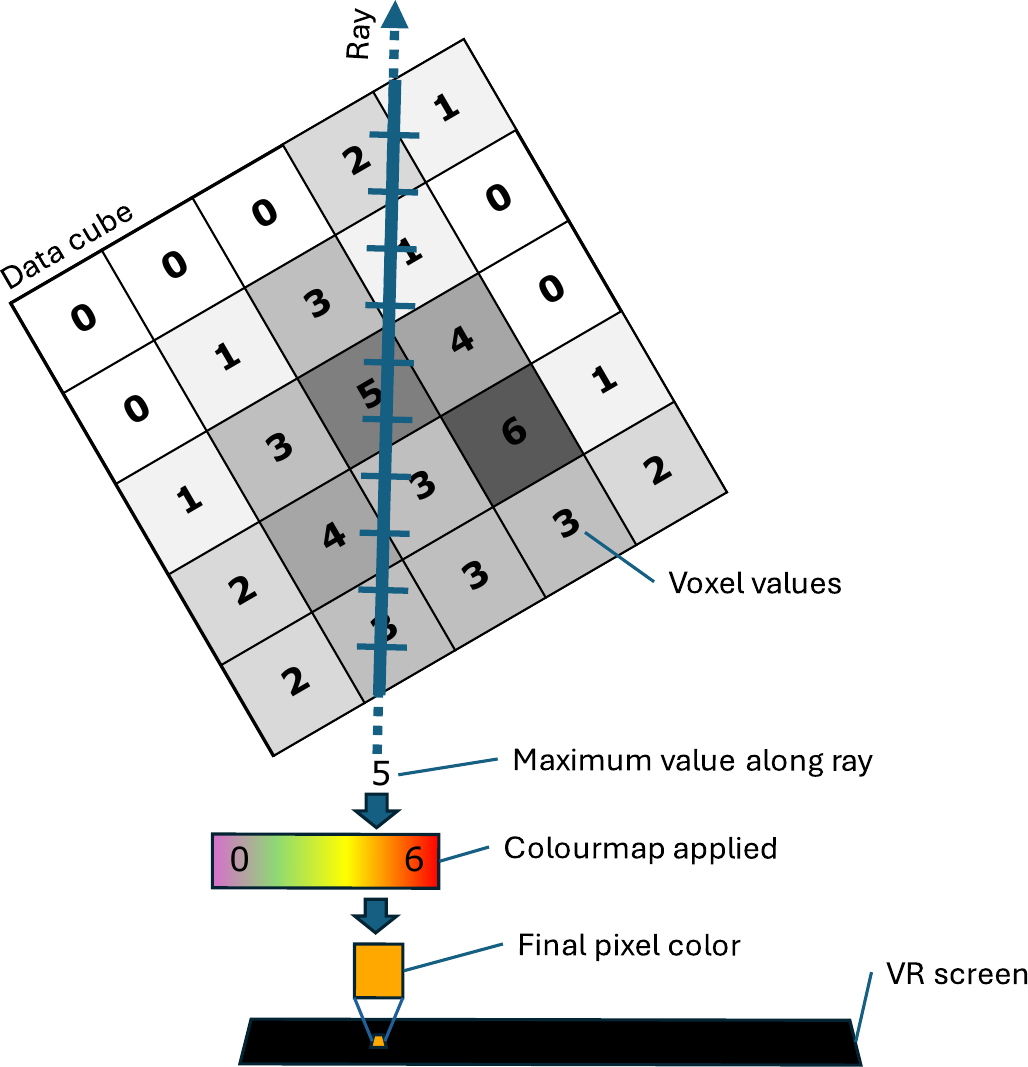}
\caption{A simplified diagram of the ray-marching algorithm used in iDaVIE. The ray is cast from the VR camera pixel and intersects the cube at two points, which are then used to calculate the interval of voxel values along the ray. The shader then applies a transform function to return the max value along the interval. A colourmap is then applied to the value to determine the colour of the pixel. The result when applied in parallel over the whole of two VR screens is a 3D rendering of the data cube voxels with a stereoscopic effect.}
\label{fig:ray_marching}
\end{figure}

In order to render the data cube in the scene, iDaVIE uses our custom shader implemented as a material attached to the cube object. The shader is written in HLSL (High-Level Shading Language) and is designed to work with Unity's Built-In Render Pipeline. The shader uses a ray-marching algorithm to render the data as a cube of voxels, with each voxel representing a single data point in the cube. Figure~\ref{fig:ray_marching} shows a simplified version of the ray-marching algorithm used in iDaVIE. 

The shader makes use of the Ray-Box Intersection algorithm \cite{haines-raytracing} which simplifies the ray calculation by treating the cube as six planes representing the six faces of the cube. By applying booleans of whether the ray is entering or exiting slabs in the x,y, and z directions, the shader can quickly calculate the interval that lies within the box. A transform function then returns a designated value along the interval. The default transform we use is the maximum value, which is useful in the context of astronomy as the ``interesting'' regions are usually the higher signal regions. We also have the option to return the average along the interval, which has shown the possibility of recognizing regions of absorption, i.e. negative voxel values (see Section~\ref{rendering_enhancements}). After the max (or mean) value is calculated, the fragment shader then applies a colourmap to the value. The colourmap is a 1D slice of a 2D texture that is sampled based on the value of the voxel, normalised between 0 and 1. The 2D colour texture is an adaption of a selected colourmap from Matplotlib \cite{hunter2007matplotlib}. The corresponding projected pixel colour is then blended with a linearly increasing alpha value from 0 to 1 normalised to the data values. This is done to give a more transparent effect to the lower values, which is useful for visualising the data cube as a whole, especially in the context of astronomy where high signal regions are of most interest. The result, when done individually for both eye screens of the VR headset, results in a stereoscopic effect that gives the illusion of depth for voxels in the data cube.

In addition to the colour and alpha transparency being applied, the shader also checks when in \texttt{Creating Selection} state if the voxel of interest along the line of sight sits within the user's selection box. If it does, the colour is applied at full saturation, otherwise it is at a lower saturation. This provides a visual cue to the user that the voxel is part of the selection box, allowing them to see which voxels they are selecting in the scene.

When using masks with the data cube, the shader simultaneously checks the mask 3D texture when accumulating along the ray, depending on the mask mode. In ``Apply Mask'' mode, the shader does not read values along the ray if the mask value is zero, effectively ignoring those data voxels. In the ``Inverse Mask'' mode, the shader only reads values along the ray if the mask value is zero, effectively ignoring the masked voxels to expose the residual data voxels. Lastly, in ``Mask Isolate'' mode, the shader treats the 3D mask texture as a normal data cube, but applies only a single colour if the voxel value is non-zero.

Because datacubes can potentially be on the order of 100's of gigabytes in size, with multiple terabytes in the future, we needed an effective way to balance flexibility regarding acceptable cube size to the strict performance requirements of virtual reality. Specifically, a framerate above 90 frames per second must be maintained to avoid discomfort for the user \cite{meta2025vrguidelines}. This is accomplished by downsampling the cube via an algorithm that condenses groups of voxels to single larger ones with a chosen window size and average or maximum value filter based on a config option (see Section~\ref{parsing_data}). Users can adjust this cube limit along with sampling steps, but extreme care must be taken to avoid framerate drops to avoid user discomfort. Additionally, a static foveated rendering can be toggled that reduces the sampling steps on the outer edges of the display. For an eye-tracking headset, this would eventually follow the centre of vision, so the maximum step size could be achieved in the foveated centre.

In early tests \cite{sivitilli_adass} it was found that there was a clear dip in performance when the headset view pointed down the y and z axes. Due to the way the ray-marching algorithm is implemented and the way the data is stored in the 3D texture, fewer cache hits occurred compared to when the user looked along the x-axis. This was important to document as a user could think their hardware was capable of higher cube sizes when testing in one view or a skinny cube that allow cache hits along the x-direction. Another key finding was how the 3D texture is sampled. While bilinear filtering is the default for 3D textures in Unity, and would be expected to give a desired, smooth effect to the rendering, it was found that this should be turned off. The ``blocky'' nature of the rendered cloud gives a more distinct indication of the different voxel values. This is especially important in the context of astronomy where the user needs to be able to clearly differentiate between voxel values.

\subsection{Feature System}
\label{feature_system}
\begin{figure}
  \centering
  \includegraphics[width=1.0\textwidth]{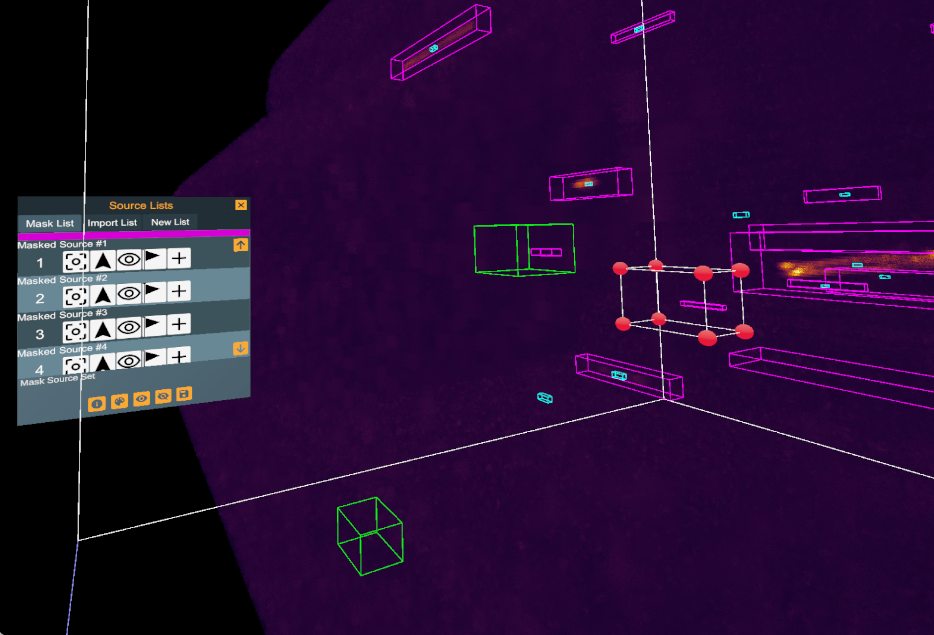}
  \caption{Here the various types of Features are overlaying the data cube. The magenta boxes are generated by bounding boxes for the loaded mask file. The Source Lists menu in the background displays the resulting list of Masked Features. The small cyan boxes are Imported Features from an external VOTable catalogue. The green boxes were added to the New Features list. The selection box (which is a special, temporary New Feature) is shown as a white box with intractable anchor points on the corners used to adjust its size.}
  \label{fig:feature_system}
  \end{figure}

In volumetric data, astronomers are often required to locate and interact with specific regions of interest. These can be from a masked cluster of voxels indicating a galaxy or simply positions in space for comparison with previous astronomical catalogues. In order to allow the user to mark and interrogate these regions in the data cube, iDaVIE implements a Feature system (denoted with a capital F). A Feature is a glowing cube outline overlaying the data cube that indicates a region of interest. In iDaVIE these come in three flavours: Masked Features, Imported Features, and New (or User-Defined) Features. All of these can be accessed through the Feature Menu in the VR environment (see Figure~\ref{fig:feature_system}).

Masked Features are generated from the mask file loaded into the scene. The \texttt{DataAnalysis} plug-in calculates the number of mask regions (voxels with non-zero mask values) in the cube and creates a list of Features with the corresponding mask region number and bounds. These are displayed in the Mask List tab of the Feature Menu. The user can select a Masked Feature to view its properties, including its bounds, mask region number, and other realtime calculated statistics (e.g., voxel count, total and peak flux, flux-weighted centroid, spectral profile, and line width $W_{20}$).

Imported Features are generated from one or more catalogues imported using the SOURCES tab of the desktop GUI. The user can select a catalogue file to load that is stored locally on the PC. The catalogue file in VOTable format \cite{VOTable_definition} is parsed and a list of Features is created with the corresponding coordinates and properties. These are displayed in the Imported List tab of the Feature Menu. The user can select an Imported Feature to view its properties, including its coordinates and other column data that the user toggled to import.

New (or User-Defined) Features are created by the user in the VR environment. The user can create a new Feature by selecting a region of interest in the data cube using the box selection tool. The user can then adjust the size and position of the Feature using the controller. This Feature can then be saved to the New List tab of the Feature Menu. This can also be done to Features generated from the Masked or Imported lists. The user can also delete a Feature from the New List tab by selecting it and pressing the delete button. If desired, the user can also export the New Features to a catalogue file in VOTable format. This allows the user to save their own catalogues of regions of interest in the data cube for later use or sharing with other astronomers.

\section{Use Cases}

There are several early published scientific uses that have demonstrated the benefit of incorporating iDaVIE into the analysis pipeline; here, we report some of these examples as a representative set, but more are available in the literature. The first involved the analysis of MeerKAT \cite{meerkat_paper} neutral hydrogen observations of the Fornax A group \cite{kleiner2021}. iDaVIE was used to inspect and refine the detection mask for neutral hydrogen sources in the galaxy group. Researchers imported the HI cube in the VR environment, overlaid the detection mask generated from the SoFIA source-finding algorithm and then adjusted the mask using iDaVIE's interaction tools. This allowed them to either identify real HI emission that was missed or remove spurious detections, ensuring a more accurate dataset. By leveraging iDaVIE, the team was able to interact with the data in a 3D immersive setting, improving their ability to distinguish between noise and actual HI sources. This process led to the inclusion of two additional sources in the detection mask that were originally excluded due to reliability thresholds.

Similarly, in a MeerKAT survey of the Fornax cluster \cite{serra2023}, iDaVIE was used to verify real HI emission from SoFIA on various angular and velocity scales of a cube through visual inspection. This effectively guided the creation of a ``clean mask'' to reveal the first evidence of ram pressure shaping the distribution of HI in a galaxy in the Fornax cluster. 

In yet another MeerKAT study \cite{marcin2024}, researchers conducted a 2.3-hour observation and serendipitously discovered 49 HI-rich galaxies, many of which were members of at least three distinct galaxy groups, potentially part of a larger supergroup structure. Several of these galaxies showed signs of interaction through disturbed HI morphologies and velocity fields, and some were found to have unusually high star formation rates or gas content. To help identify and analyse these features, the team used iDaVIE, which allowed them to explore the HI data cubes spatially and spectrally. iDaVIE enabled efficient visual inspection of the data, supporting the confirmation of real HI emission features and the recognition of extended structures and interactions between galaxies—insights that complemented and enhanced results from automated source-finding algorithms. 

Finally, another group \cite{Deg_2023} discovered two polar ring galaxies in HI data from the Widefield ASKAP L-band Legacy All-sky Blind surveY (WALLABY) \cite{wallaby_paper} using iDaVIE. With the mask painting tool, they were able to isolate anomalous gas detections from the main galactic disks, thus identifying the polar ring features for follow-up kinematic modelling. Specifically, the authors note the advantage that iDaVIE has for low signal-to-noise observations, where the entire data cube can be visible while interaction can be done at the individual voxel-level.

\section{Benchmarking and Performance}
\label{sec:performance}

Performance benchmarks were conducted on a Windows 11 system equipped with an Intel Core Ultra 9 185H CPU (16 cores), 32GB DDR5 RAM, and an NVIDIA RTX 4070 Laptop GPU. The VR headset (Meta Quest 3) was connected via a wired link to avoid network-induced variability. The headset refresh rate was configured to 90Hz during testing. Measurements were obtained using a standalone build of iDaVIE.

Synthetic data cubes of increasing size ($256^3$, $512^3$, $1024^3$, and $1536^3$ voxels), together with corresponding mask cubes, were generated to evaluate scaling behaviour. In addition, a representative science cube (1046$\times$952$\times$398 voxels) containing HI survey data of the Fornax galaxy cluster \cite{Fornax_paper}, along with an accompanying mask generated using the SoFiA source-finding pipeline, was included to assess performance on real observational data. For each dataset, three independent trials were performed, and the reported values correspond to their mean.

Load time was measured from user initiation of the loading process to full renderer initialisation when user VR interactivity would normally begin. Peak private memory usage was recorded relative to the baseline process memory at load start, thereby isolating the additional memory consumption attributable to cube loading and texture preparation. VRAM usage was computed from the final downsampled texture dimensions and voxel precision (4 bytes for data cubes, 2 bytes for mask cubes).

For framerate evaluation, a scripted interaction was employed. The cube was temporarily attached to the headset camera and translated in camera-local coordinates from a fixed starting distance of 2\,m to the cube centre (0\,m). The framerate was sampled once per rendered frame during this single continuous translation phase. Median, minimum, and mean FPS values were computed from these samples. The benchmarking script implementing this protocol is included in the public iDaVIE repository for reproducibility.

Results are summarised in Tables~\ref{tab:loading-performance} and~\ref{tab:fps-performance}. Cube initialisation and memory usage are shown separately from runtime framerate metrics to distinguish dataset preparation costs from interactive rendering performance. Load time and peak RAM usage increased predictably with dataset size, reaching approximately 22\,GB of peak private memory for the largest synthetic cube (native $1536^3$). For the same cube, which was downsampled to an effective texture size of $384\times384\times512$, median framerates remained close to the headset refresh rate ($\sim$90\,Hz), indicating stable interactive performance under the scripted inspection protocol. Despite the substantial growth in native voxel count, framerate remains largely stable across configurations, suggesting that the adaptive downsampling strategy helps maintain interactive performance. The primary impact of larger datasets is therefore observed in increased loading time and memory consumption rather than runtime rendering performance. The representative science cube demonstrated performance comparable to the mid-range synthetic cases, confirming practical usability for typical observational datasets.

\setlength{\tabcolsep}{4pt}
\begin{table}[htbp]
\centering
\caption{Loading and memory performance averaged over three runs. 
``Native Dim.'' denotes the original cube dimensions prior to downsampling. 
``Down'' indicates the integer downsampling factors applied along each axis ($x\times y\times z$). 
RAM denotes peak private memory usage during cube initialisation, and VRAM corresponds to the memory footprint of the resulting 3D textures.}
\label{tab:loading-performance}
\begin{tabular}{lccccc}
\hline
Cube & Native Dim. & Down & Load (s) & RAM (MB) & VRAM (MB) \\
\hline
Synthetic S & $256^3$  & 1×1×1 & 1.09 & 820  & 96  \\
Synthetic M & $512^3$  & 2×2×1 & 2.05 & 1449 & 192 \\
Synthetic L & $1024^3$ & 3×3×2 & 10.34 & 7240 & 341 \\
Synthetic XL & $1536^3$ & 4×4×3 & 45.37 & 22040 & 432 \\
Science     & 1046×952×398 & 3×3×1 & 2.86 & 3151 & 251 \\
\hline
\end{tabular}
\end{table}

\begin{table}[htbp]
\centering
\caption{Framerate statistics recorded during the scripted cube translation protocol. FPS$_{\mathrm{med}}$, FPS$_{\mathrm{min}}$, and FPS$_{\mathrm{mean}}$ denote the median, minimum, and arithmetic mean framerate measured over the sampling interval. Values are averaged over three runs.}
\label{tab:fps-performance}
\begin{tabular}{lccc}
\hline
Cube & FPS$_{\mathrm{med}}$ & FPS$_{\mathrm{min}}$ & FPS$_{\mathrm{mean}}$ \\
\hline
Synthetic S & 90.5 & 38.3 & 89.5 \\
Synthetic M & 89.3 & 35.2 & 85.7 \\
Synthetic L & 88.7 & 39.1 & 81.9 \\
Synthetic XL & 88.9 & 35.0 & 83.9 \\
Science     & 90.6 & 42.3 & 90.1 \\
\hline
\end{tabular}
\end{table}

\section{Availability}
\label{label_availability}
The iDaVIE software is publicly available on GitHub at \url{https://github.com/idia-astro/iDaVIE}. A DOI for version 1.0 of the software has been registered via Zenodo: \url{https://doi.org/10.5281/zenodo.13752029}~\citep{idavie_zenodo}. Further documentation, including installation instructions, user guides, and developer resources, can be found on the documentation site at \url{https://idavie.readthedocs.io/}.

The software is distributed under the open-source GNU Lesser General Public Licence version 3, allowing for free use, modification, and distribution. iDaVIE is currently supported on a VR-ready Windows PC with a dedicated GPU and a compatible VR headset (e.g., HTC Vive, Valve Index, or Meta Quest series). In addition, the free SteamVR software platform is required to be installed through the Steam digital distribution service application to enable VR headset connectivity with the PC.

Installation of the software currently only requires downloading and extracting the \texttt{.zip} file from the GitHub repository. The software is self-contained and does not require any additional dependencies or libraries to be installed. The user can then run the \texttt{iDaVIE.exe} file to launch the software.

As the project is open-source, the source code is available for anyone to view, build, modify, and contribute to. The project is actively maintained and updated with new features and improvements based on user feedback and contributions from the community. Users are encouraged to report any bugs or issues they encounter while using the software via the GitHub issue page. Building the software requires the Unity game engine and the necessary external libraries and plug-ins to be installed. Details on building the software from source are provided in the GitHub repository.

\section{Challenges and Future Work}

The release of iDaVIE version 1.0 marks a significant milestone, but there are several challenges and opportunities for future development to enhance its functionality and usability. Below, we outline key areas of focus going forward with the project. Where noted, these features will be part of a release candidate for version 1.1, under active development at the time of writing.

\subsection{Subcube Loading and HDU Selection}
Currently, users can only load a full 3D FITS cube with a single HDU. If the cube is large, iDaVIE is able to downsample the voxels to make the VR session manageable. However, for very large cubes that are either a significant portion of system RAM or well above the 4GB Unity texture limit, this can lead to unbearable load times and also starts to cause the ``washing out'' of features in the data prior to cropping. Memory usage scales with the cube and mask size (see Section~\ref{sec:performance}); practical use therefore requires sufficient system RAM to hold the full-resolution cube (and mask) alongside the downsampled textures. If RAM is insufficient, the operating system may page to disk, resulting in substantially longer load times.

A major planned enhancement is the ability to load a subcube, allowing users to specify bounds for a contiguous portion of the data cube. This feature will require a significant rework of file operations. Additionally, support for selecting different HDUs (Header Data Units) will be introduced, enabling compatibility with instruments like MUSE, NIRSpec, and MIRI, which often store data in the second HDU. These features will be developed concurrently as their functionality is closely related. This is being targeted for release in version 1.1.

\subsection{Scripting and Video Generation}
The predominant method to share visuals from iDaVIE is through screenshots or screen recording of the VR user's view. This is usually sufficient for still images, but to demonstrate or present more complex features in the data, video is necessary to reveal 3D cues. However, the natural moving headset of the user can make such a recording unsettling or even unwatchable on a flatscreen. In the past we have been able to make custom videos by hardcoding scripted behaviour of a second camera in the Unity scene, however this is generally labour-intensive. To improve the quality of recorded visualisations, a scripting language or API is being developed. This will allow users to create smooth, high-quality, reproducible videos by controlling a Unity camera's movement through the data. This is also being targeted for release in version 1.1.

\subsection{Rendering Enhancements}
\label{rendering_enhancements}
iDaVIE's rendering mostly caters to users looking for bright regions in data cubes due to the default maximum-value voxel transfer function. Absorption features in the data therefore disappear, even when statistically significant as they are ignored in favour of pure-noise positive voxels. Although the mean transfer function for the ray-marching shader somewhat allows otherwise, other methods are needed to represent more complex features in the voxels. A way to solve this issue would be to implement a hybrid algorithm which shows maximum or minimum values along the line of sight, depending on which is larger in absolute value. This could lead to an even more advanced shader based on radiative transfer equations to account for foreground absorption, expanding the tool's applicability to a broader range of datasets. 

\subsection{Separation of Visualisation and Analysis Tools}
The visualisation and analysis tools in the native plug-ins are tightly coupled at the moment. Future work will focus on separating these components, allowing analysis tools to be implemented as third-party plug-ins. This modular approach will enable multidisciplinary use, where users can download iDaVIE as a visualisation tool and add analysis plug-ins relevant to their field.

\subsection{Particle Dataset Visualisation}
Expanding iDaVIE to support sparse multi-parameter datasets, such as those from simulations or survey catalogues, is a medium-term goal. A prototype for visualising particle datasets has already been created, and sample scenes are available in the source code. Further development will make this feature publicly available as stand-alone component of the software with its own dedicated GUI and interaction tools. Although not yet publicly available, readers are welcome to request experimental builds if they would like to take part in early beta-testing.

\subsection{Long-Term Goals}
Other long-term development goals for iDaVIE include:
\begin{itemize}
    \item Remote streaming of large datasets to local client VR headsets
    \item Integrating a Python console for advanced scripting and analysis
    \item Adding additional visualisation modes for regions of interest, such as iso-contours/surfaces
    \item Creating a multiplayer or spectator mode for collaborative analysis, including a digital planetarium streaming mode for immersive presentations \cite{faherty2020}
    \item Integrating a Virtual Observatory (VO) system for dynamic retrieval of images from data catalogues
    \item Supporting multiple cubes or time-series visualisation, which will require significant performance optimisations
    \item Implementing state- or workspace-saving to enhance user workflow integration and collaboration
\end{itemize}

These planned developments aim to make iDaVIE a more versatile and powerful tool for immersive data visualisation and analysis, catering to the evolving needs of the scientific community.

\section{Conclusion}

iDaVIE began as a proof of concept for visualising large 3D data cubes in virtual reality but has since evolved into a fully functional, open-source software package that integrates immersive visualisation with practical data interaction tools. By addressing the challenges of navigating multidimensional astronomical datasets, iDaVIE offers researchers an intuitive environment where the cognitive overhead of data manipulation is intended to be reduced, allowing them to focus on scientific discovery.

The software's key innovations for immersive visualisation, such as native FITS file support, interactive selection and masking tools, real-time quantitative analysis, and catalogue integration, have demonstrated clear advantages in scientific workflows. Published use cases with MeerKAT and ASKAP data have shown how iDaVIE enables astronomers to verify automated pipeline results, refine detection masks, and discover previously overlooked features in HI data cubes. These applications validate our approach of complementing rather than replacing traditional analysis tools.

As astronomy continues to generate increasingly large and complex datasets, tools like iDaVIE become essential bridges between automated processing pipelines and human guidance. The immersive, spatial interaction offered by VR is expected to reduce the translation burden between 3D data and 2D interfaces, providing a more direct connection to multidimensional data. This is particularly valuable for quality control, source verification, and complex feature identification, tasks where human pattern recognition excels but traditional interfaces impose limitations.

By making iDaVIE freely available under an open-source licence, we aim to foster community adoption and contribution, allowing the software to grow alongside evolving research needs. While version 1.0 represents a significant milestone, the planned enhancements including subcube loading, advanced rendering modes, scripting capabilities, and collaborative features will further extend its utility across astronomy and potentially other scientific disciplines.

iDaVIE demonstrates how immersive technologies can be meaningfully integrated into scientific workflows, offering researchers new perspectives on their data while maintaining the rigour and reproducibility required for professional research. As virtual reality hardware becomes more accessible and powerful, tools like iDaVIE will play an increasingly important role in helping scientists navigate the challenges of big data in their respective fields.

\section*{CRediT authorship contribution statement}

\textbf{Alexander Sivitilli:} Conceptualization, Methodology, Software, Validation, Writing – Original Draft.
\textbf{Lucia Marchetti:} Conceptualization, Resources, Supervision, Project Administration, Funding acquisition, Writing – Review \& Editing.
\textbf{Angus Comrie:} Conceptualization, Methodology, Software, Supervision.
\textbf{P. Cilliers Pretorius:} Methodology, Software, Writing – Review \& Editing.
\textbf{Thijs (J.M.) van der Hulst:} Validation, Data Curation, Supervision, Writing – Review \& Editing.
\textbf{Fabio Vitello:} Methodology, Software.
\textbf{D.J. Pisano:} Data Curation, Supervision, Funding acquisition, Writing – Review \& Editing.
\textbf{Ugo Becciani:} Supervision, Funding acquisition.
\textbf{A. Russell Taylor:} Conceptualization, Resources, Supervision, Funding acquisition.
\textbf{Paolo Serra:} Conceptualization, Validation, Data Curation.
\textbf{Mayhew Steyn:} Methodology, Software.
\textbf{Michaela van Zyl:} Methodology, Software.

\section*{Acknowledgements}
We would like to acknowledge the essential contributions of the iDaVIE development team and the valuable feedback from our user community. In particular, Dr Marcin Glowacki has been an key tester from the early days and Professor Neal Katz has provided critical guidance for the upcoming particle rendering features. A special mention also goes out to Dr Michelle Cluver for her important role in the project's direction, including the suggestion of its final name. In addition, we thank Professor Sally Macfarlane for designing the project logo in and for reliable project advice at multiple stages.

We also extend our heartfelt gratitude to Professor Thomas Jarrett for his invaluable support and guidance throughout the project. His expertise and insights were instrumental in shaping iDaVIE into the tool it is today. He will be sorely missed by the team and the wider astronomical community.

An extra special thank you goes to IDIA and the University of Cape Town for providing the equipment, resources, and facilities necessary for the development and testing of iDaVIE.

The iDaVIE team greatly acknowledge the support from the South African Research Chairs Initiative of the Department of Science and Technology and National Research Foundation.

This project has received funding from the European Research Council (ERC) under the European Union's Seventh Framework Programme (FP/2007-2013)/ERC (grant agreement no. 291531, “HIStoryNU”) and the Horizon 2020 research and innovation programme (grant agreement no. 679627, ``FORNAX''), from MAECI (Italian Ministry of Foreign Affairs and International Cooperation) Grant Numbers PGR ZA23GR03 (RADIOMAP) and PGR ZA18GR02 (RADIOSKY2020), from DSTNRF Grant Number 113121 as part of the ISARP Joint Research Scheme, from ISARP RADIOMAP Joint Research Scheme (DSI-NRF Grant Number 150551) and from the CPRR Projects (DSI-NRF Grant Number SRUG2204254729).

Figures throughout the paper make use of a datacube from the Australia Telescope Compact Array (ATCA) HI survey of the Fornax galaxy cluster \cite{Fornax_paper}.


\section*{Glossary}

\begin{itemize}
    \item APERTIF: APERture Tile In Focus, a phased array feed system on the Westerbork Synthesis Radio Telescope.
    \item ASKAP: Australian Square Kilometre Array Pathfinder, a radio telescope array in Australia.
    \item CARTA: Cube Analysis and Rendering Tool for Astronomy, a browser-based desktop application for visualising large data sets.
    \item Downsampling: Reducing resolution by aggregating neighbouring voxels (e.g., mean or max) to reduce the rendering cost of a region of the cube.
    \item FITS: Flexible Image Transport System, the standard file format for astronomical data (images, spectra, cubes).
    \item Feature: A region of interest in the data cube, typically marked for selection, analysis, or annotation.
    \item GPU: Graphics Processing Unit, hardware that accelerates rendering and parallel computation.
    \item HDU: Header Data Unit, a single header+data block in a FITS file; files can contain multiple HDUs.
    \item HI: Neutral atomic hydrogen, often studied in radio astronomy.
    \item Isosurface: A surface representing points of constant data value in a volume, often used to extract structures.
    \item LGPL: Lesser General Public License, a free software license that allows software to be freely used, modified, and distributed, with certain conditions.
    \item Mask: A secondary cube (or array) marking regions of interest with integer labels, used for selection and analysis.
    \item MeerKAT: A radio telescope array located in South Africa, used for various astrophysical observations.
    \item Moment map: A projection (e.g., zeroth, first) of a spectral cube along the spectral axis to produce summary images (integrated intensity, velocity).
    \item Rectoid: A geometric shape that is a three-dimensional analogue of a rectangle, often used in the context of data visualisation to represent volumetric data.
    \item Ray-marching: A volumetric rendering technique that marches a ray through a volume, sampling values along the ray to produce a pixel colour.
    \item S/N: Signal-to-noise ratio, a measure of the strength of an observed signal relative to background noise.
    \item SoFIA: HI Source Finding Application used for identifying astronomical sources in data cubes.
    \item Subcube: A contiguous subset of a larger data cube defined by bounds along each axis.
    \item Transfer function: A mapping from voxel values to colour and opacity used in volume rendering to emphasise features.
    \item VOTable: An XML table format commonly used in the Virtual Observatory for exchanging tabular astronomical data.
    \item VR: Virtual reality, a simulated experience that can be similar to or completely different from the real world.
    \item VRAM: Video RAM, GPU memory used to store textures and other graphics resources.
    \item Voxel: A volume element (3D pixel) representing a single data value in a data cube.
    \item WCS: World Coordinate System, metadata in FITS headers that maps pixel coordinates to sky or spectral coordinates.
\end{itemize}



\section*{Declaration of generative AI and AI-assisted technologies in the writing process}
During manuscript preparation, AS used OpenAI's GPT-4 for grammar, phrasing, and minor stylistic edits, and Google's Gemini 2.5 to assist with editing PlantUML scripts used to generate UML diagrams. The tools are not listed as authors. No confidential, personal, or unpublished data were provided to these services. All outputs were reviewed, edited, and verified by the authors, who take full responsibility for the content of this article.


\bibliographystyle{elsarticle-harv} 
\bibliography{references} 

@ARTICLE{pisano2002,
       author = {{Pisano}, D.~J. and {Wilcots}, Eric M. and {Liu}, Charles T.},
        title = "{An H I/Optical Atlas of Isolated Galaxies}",
      journal = {ApJS},
         year = 2002,
        month = oct,
       volume = {142},
       number = {2},
        pages = {161-222},
          doi = {10.1086/341787},
       adsurl = {https://ui.adsabs.harvard.edu/abs/2002ApJS..142..161P}
}

@article{lsst_paper,
  doi       = {10.3847/1538-4357/ab042c},
  url       = {https://doi.org/10.3847/1538-4357/ab042c},
  year      = {2019},
  month     = {mar},
  publisher = {The American Astronomical Society},
  volume    = {873},
  number    = {2},
  pages     = {111},
  author    = {Ivezi\'{c}, \v{Z}eljko and Kahn, Steven M. and Tyson, J. Anthony and Abel, Bob and Acosta, Emily and Allsman, Robyn and Alonso, David and AlSayyad, Yusra and Anderson, Scott F. and Andrew, John and P. Angel, James Roger and Angeli, George Z. and Ansari, Reza and Antilogus, Pierre and Araujo, Constanza and Armstrong, Robert and Arndt, Kirk T. and Astier, Pierre and Aubourg, \'{E}ric and Auza, Nicole and Axelrod, Tim S. and Bard, Deborah J. and Barr, Jeff D. and Barrau, Aurelian and Bartlett, James G. and Bauer, Amanda E. and Bauman, Brian J. and Baumont, Sylvain and Bechtol, Ellen and Bechtol, Keith and Becker, Andrew C. and Becla, Jacek and Beldica, Cristina and Bellavia, Steve and Bianco, Federica B. and Biswas, Rahul and Blanc, Guillaume and Blazek, Jonathan and Blandford, Roger D. and Bloom, Josh S. and Bogart, Joanne and Bond, Tim W. and Booth, Michael T. and Borgland, Anders W. and Borne, Kirk and Bosch, James F. and Boutigny, Dominique and Brackett, Craig A. and Bradshaw, Andrew and Brandt, William Nielsen and Brown, Michael E. and Bullock, James S. and Burchat, Patricia and Burke, David L. and Cagnoli, Gianpietro and Calabrese, Daniel and Callahan, Shawn and Callen, Alice L. and Carlin, Jeffrey L. and Carlson, Erin L. and Chandrasekharan, Srinivasan and Charles-Emerson, Glenaver and Chesley, Steve and Cheu, Elliott C. and Chiang, Hsin-Fang and Chiang, James and Chirino, Carol and Chow, Derek and Ciardi, David R. and Claver, Charles F. and Cohen-Tanugi, Johann and Cockrum, Joseph J. and Coles, Rebecca and Connolly, Andrew J. and Cook, Kem H. and Cooray, Asantha and Covey, Kevin R. and Cribbs, Chris and Cui, Wei and Cutri, Roc and Daly, Philip N. and Daniel, Scott F. and Daruich, Felipe and Daubard, Guillaume and Daues, Greg and Dawson, William and Delgado, Francisco and Dellapenna, Alfred and Peyster, Robert de and Val-Borro, Miguel de and Digel, Seth W. and Doherty, Peter and Dubois, Richard and Dubois-Felsmann, Gregory P. and Durech, Josef and Economou, Frossie and Eifler, Tim and Eracleous, Michael and Emmons, Benjamin L. and Neto, Angelo Fausti and Ferguson, Henry and Figueroa, Enrique and Fisher-Levine, Merlin and Focke, Warren and Foss, Michael D. and Frank, James and Freemon, Michael D. and Gangler, Emmanuel and Gawiser, Eric and Geary, John C. and Gee, Perry and Geha, Marla and Gessner, Charles J. B. and Gibson, Robert R. and Gilmore, D. Kirk and Glanzman, Thomas and Glick, William and Goldina, Tatiana and Goldstein, Daniel A. and Goodenow, Iain and Graham, Melissa L. and Gressler, William J. and Gris, Philippe and Guy, Leanne P. and Guyonnet, Augustin and Haller, Gunther and Harris, Ron and Hascall, Patrick A. and Haupt, Justine and Hernandez, Fabio and Herrmann, Sven and Hileman, Edward and Hoblitt, Joshua and Hodgson, John A. and Hogan, Craig and Howard, James D. and Huang, Dajun and Huffer, Michael E. and Ingraham, Patrick and Innes, Walter R. and Jacoby, Suzanne H. and Jain, Bhuvnesh and Jammes, Fabrice and Jee, M. James and Jenness, Tim and Jernigan, Garrett and Jevremovi\'{c}, Darko and Johns, Kenneth and Johnson, Anthony S. and Johnson, Margaret W. G. and Jones, R. Lynne and Juramy-Gilles, Claire and Juri\'{c}, Mario and Kalirai, Jason S. and Kallivayalil, Nitya J. and Kalmbach, Bryce and Kantor, Jeffrey P. and Karst, Pierre and Kasliwal, Mansi M. and Kelly, Heather and Kessler, Richard and Kinnison, Veronica and Kirkby, David and Knox, Lloyd and Kotov, Ivan V. and Krabbendam, Victor L. and Krughoff, K. Simon and Kub\'{a}nek, Petr and Kuczewski, John and Kulkarni, Shri and Ku, John and Kurita, Nadine R. and Lage, Craig S. and Lambert, Ron and Lange, Travis and Langton, J. Brian and Guillou, Laurent Le and Levine, Deborah and Liang, Ming and Lim, Kian-Tat and Lintott, Chris J. and Long, Kevin E. and Lopez, Margaux and Lotz, Paul J. and Lupton, Robert H. and Lust, Nate B. and MacArthur, Lauren A. and Mahabal, Ashish and Mandelbaum, Rachel and Markiewicz, Thomas W. and Marsh, Darren S. and Marshall, Philip J. and Marshall, Stuart and May, Morgan and McKercher, Robert and McQueen, Michelle and Meyers, Joshua and Migliore, Myriam and Miller, Michelle and Mills, David J. and Miraval, Connor and Moeyens, Joachim and Moolekamp, Fred E. and Monet, David G. and Moniez, Marc and Monkewitz, Serge and Montgomery, Christopher and Morrison, Christopher B. and Mueller, Fritz and Muller, Gary P. and Arancibia, Freddy Mu\~{n}oz and Neill, Douglas R. and Newbry, Scott P. and Nief, Jean-Yves and Nomerotski, Andrei and Nordby, Martin and O'Connor, Paul and Oliver, John and Olivier, Scot S. and Olsen, Knut and O'Mullane, William and Ortiz, Sandra and Osier, Shawn and Owen, Russell E. and Pain, Reynald and Palecek, Paul E. and Parejko, John K. and Parsons, James B. and Pease, Nathan M. and Peterson, J. Matt and Peterson, John R. and Petravick, Donald L. and Petrick, M. E. Libby and Petry, Cathy E. and Pierfederici, Francesco and Pietrowicz, Stephen and Pike, Rob and Pinto, Philip A. and Plante, Raymond and Plate, Stephen and Plutchak, Joel P. and Price, Paul A. and Prouza, Michael and Radeka, Veljko and Rajagopal, Jayadev and Rasmussen, Andrew P. and Regnault, Nicolas and Reil, Kevin A. and Reiss, David J. and Reuter, Michael A. and Ridgway, Stephen T. and Riot, Vincent J. and Ritz, Steve and Robinson, Sean and Roby, William and Roodman, Aaron and Rosing, Wayne and Roucelle, Cecille and Rumore, Matthew R. and Russo, Stefano and Saha, Abhijit and Sassolas, Benoit and Schalk, Terry L. and Schellart, Pim and Schindler, Rafe H. and Schmidt, Samuel and Schneider, Donald P. and Schneider, Michael D. and Schoening, William and Schumacher, German and Schwamb, Megan E. and Sebag, Jacques and Selvy, Brian and Sembroski, Glenn H. and Seppala, Lynn G. and Serio, Andrew and Serrano, Eduardo and Shaw, Richard A. and Shipsey, Ian and Sick, Jonathan and Silvestri, Nicole and Slater, Colin T. and Smith, J. Allyn and Smith, R. Chris and Sobhani, Shahram and Soldahl, Christine and Storrie-Lombardi, Lisa and Stover, Edward and Strauss, Michael A. and Street, Rachel A. and Stubbs, Christopher W. and Sullivan, Ian S. and Sweeney, Donald and Swinbank, John D. and Szalay, Alexander and Takacs, Peter and Tether, Stephen A. and Thaler, Jon J. and Thayer, John Gregg and Thomas, Sandrine and Thornton, Adam J. and Thukral, Vaikunth and Tice, Jeffrey and Trilling, David E. and Turri, Max and Berg, Richard Van and Berk, Daniel Vanden and Vetter, Kurt and Virieux, Francoise and Vucina, Tomislav and Wahl, William and Walkowicz, Lucianne and Walsh, Brian and Walter, Christopher W. and Wang, Daniel L. and Wang, Shin-Yawn and Warner, Michael and Wiecha, Oliver and Willman, Beth and Winters, Scott E. and Wittman, David and Wolff, Sidney C. and Wood-Vasey, W. Michael and Wu, Xiuqin and Xin, Bo and Yoachim, Peter and Zhan, Hu},
  title     = {LSST: From Science Drivers to Reference Design and Anticipated Data Products},
  journal   = {The Astrophysical Journal}
}

@article{ska_data_paper,
  author   = {Dewdney, Peter E. and Hall, Peter J. and Schilizzi, Richard T. and Lazio, T. Joseph L. W.},
  journal  = {Proceedings of the IEEE},
  title    = {The Square Kilometre Array},
  year     = {2009},
  volume   = {97},
  number   = {8},
  pages    = {1482-1496},
  doi      = {10.1109/JPROC.2009.2021005}
}

@article{sweller_clt,
  author   = {Sweller, John},
  title    = {Cognitive Load During Problem Solving: Effects on Learning},
  journal  = {Cognitive Science},
  volume   = {12},
  number   = {2},
  pages    = {257-285},
  doi      = {10.1207/s15516709cog1202_4},
  url      = {https://onlinelibrary.wiley.com/doi/abs/10.1207/s15516709cog1202_4},
  year     = {1988}
}

@article{juliano2022increased,
  title     = {Increased cognitive load in immersive virtual reality during visuomotor adaptation is associated with decreased long-term retention and context transfer},
  author    = {Juliano, Julia M and Schweighofer, Nicolas and Liew, Sook-Lei},
  journal   = {Journal of NeuroEngineering and Rehabilitation},
  volume    = {19},
  number    = {1},
  pages     = {106},
  year      = {2022},
  publisher = {Springer}
}

@article{bigdata_paper,
  author   = {Zhang, Yanxia and Zhao, Yongheng},
  doi      = {10.5334/dsj-2015-011},
  journal  = {Data Science Journal},
  month    = {May},
  title    = {Astronomy in the Big Data Era},
  year     = {2015}
}

@ARTICLE{GASS,
       author = {{McClure-Griffiths}, N.~M. and {Pisano}, D.~J. and {Calabretta}, M.~R. and {Ford}, H. Alyson and {Lockman}, Felix J. and {Staveley-Smith}, L. and {Kalberla}, P.~M.~W. and {Bailin}, J. and {Dedes}, L. and {Janowiecki}, S. and {Gibson}, B.~K. and {Murphy}, T. and {Nakanishi}, H. and {Newton-McGee}, K.},
        title = "{Gass: The Parkes Galactic All-Sky Survey. I. Survey Description, Goals, and Initial Data Release}",
      journal = {ApJS},
         year = 2009,
        month = apr,
       volume = {181},
       number = {2},
        pages = {398-412},
          doi = {10.1088/0067-0049/181/2/398},
archivePrefix = {arXiv},
       eprint = {0901.1159},
 primaryClass = {astro-ph.GA},
       adsurl = {https://ui.adsabs.harvard.edu/abs/2009ApJS..181..398M}
}

@ARTICLE{HI4PI,
       author = {{HI4PI Collaboration} and {Ben Bekhti}, N. and {Fl{\"o}er}, L. and {Keller}, R. and {Kerp}, J. and {Lenz}, D. and {Winkel}, B. and {Bailin}, J. and {Calabretta}, M.~R. and {Dedes}, L. and {Ford}, H.~A. and {Gibson}, B.~K. and {Haud}, U. and {Janowiecki}, S. and {Kalberla}, P.~M.~W. and {Lockman}, F.~J. and {McClure-Griffiths}, N.~M. and {Murphy}, T. and {Nakanishi}, H. and {Pisano}, D.~J. and {Staveley-Smith}, L.},
        title = "{HI4PI: A full-sky H I survey based on EBHIS and GASS}",
      journal = {A\&A},
         year = 2016,
        month = oct,
       volume = {594},
          eid = {A116},
        pages = {A116},
          doi = {10.1051/0004-6361/201629178},
archivePrefix = {arXiv},
       eprint = {1610.06175},
 primaryClass = {astro-ph.GA},
       adsurl = {https://ui.adsabs.harvard.edu/abs/2016A&A...594A.116H}
}

@ARTICLE{CHILES,
       author = {{Luber}, Nicholas and {Pisano}, D.~J. and {van Gorkom}, J.~H. and {Blue Bird}, Julia and {Dodson}, Richard and {Gim}, Hansung B. and {Hess}, Kelley M. and {Hunt}, Lucas R. and {Lucero}, Danielle and {Meyer}, Martin and {Momjian}, Emmanuel and {Yun}, Min S.},
        title = "{CHILES VIII: Probing Evolution of Average HI Content in Star Forming Galaxies over the Past 5 Billion Years}",
      journal = {arXiv e-prints},
         year = 2025,
        month = apr,
          eid = {arXiv:2504.02100},
        pages = {arXiv:2504.02100},
          doi = {10.48550/arXiv.2504.02100},
archivePrefix = {arXiv},
       eprint = {2504.02100},
 primaryClass = {astro-ph.GA},
       adsurl = {https://ui.adsabs.harvard.edu/abs/2025arXiv250402100L}
}

@INPROCEEDINGS{LADUMA,
       author = {{Blyth}, S. and {Baker}, A.~J. and {Holwerda}, B. and {Bouchard}, A. and {Catinella}, B. and {Chemin}, L. and {Cunnama}, D. and {Dav{\'e}}, R. and {Faltenbacher}, A. and {February}, S. and {Fern{\'a}ndez}, X. and {Gawiser}, E. and {Heywood}, I. and {Kere{\v{s}}}, D. and {Kl{\"o}ckner}, H.~R. and {Lah}, P. and {Lochner}, M. and {Maddox}, N. and {Makhathini}, S. and {Moodley}, K. and {Morganti}, R. and {Obreschkow}, D. and {Oh}, S.~H. and {Pisano}, D.~J. and {Popping}, A. and {Popping}, G. and {Ravindranath}, S. and {Schinnerer}, E. and {Sheth}, K. and {Skelton}, R. and {Smith}, M. and {Srianand}, R. and {Staveley-Smith}, L. and {Vaccari}, M. and {Vaisanen}, P. and {Walter}, F. and {Rawlings}, S. and {Bassett}, B.~A. and {Bershady}, M.~A. and {Briggs}, F.~H. and {Crawford}, S.~M. and {Cress}, C.~M. and {Darling}, J.~K. and {Deane}, R.~P. and {de Blok}, G. and {Elson}, E.~C. and {Frank}, B.~S. and {Henning}, P.~A. and {Hess}, K.~M. and {Hughes}, J.~P. and {Jarvis}, M.~J. and {Kannappan}, S.~J. and {Katz}, N.~S. and {Kraan-Korteweg}, R.~C. and {Lehnert}, M.~D. and {Leroy}, A.~K. and {Meurer}, G.~R. and {Meyer}, M.~J. and {Pisano}, D.~J. and {Schr{\"o}der}, A.~C. and {Smirnov}, O.~M. and {Somerville}, R.~S. and {Stewart}, I.~M. and {van der Heyden}, K.~J. and {Verheijen}, M.~A.~W. and {Wilcots}, E.~M. and {Williams}, T.~B. and {Woudt}, P.~A. and {Wu}, J.~F. and {Zwaan}, M.~A. and {Zwart}, J.~T.~L. and {Oosterloo}, T.~A. and {van Drie}, W.},
        title = "{LADUMA: Looking at the Distant Universe with the MeerKAT Array}",
    booktitle = {MeerKAT Science: On the Pathway to the SKA},
         year = 2016,
        month = jan,
          eid = {4},
        pages = {4},
          doi = {10.22323/1.277.0004},
       adsurl = {https://ui.adsabs.harvard.edu/abs/2016mks..confE...4B}
}

@article{kleiner2021,
  author = {Kleiner, D. and Serra, P. and Maccagni, F. M. and Venhola, A. and Morokuma-Matsui, K. and Peletier, R. and Iodice, E. and Raj, M. A. and de Blok, W. J. G. and Comrie, A. and J{\'o}zsa, G. I. G. and Kamphuis, P. and Loni, A. and Loubser, S. I. and Moln{\'a}r, D. Cs. and Passmoor, S. S. and Ramatsoku, M. and Sivitilli, A. and Smirnov, O. and Thorat, K. and Vitello, F.},
  title = {A MeerKAT view of pre-processing in the Fornax A group},
  journal = {A\&A},
  year = {2021},
  volume = {648},
  pages = {A32},
  doi = {10.1051/0004-6361/202039898},
  url = {https://doi.org/10.1051/0004-6361/202039898}
}

@article{wells1981fits,
  author    = {Wells, D. C. and Greisen, E. W. and Harten, R. H.},
  title     = {FITS - a Flexible Image Transport System},
  journal   = {Astronomy and Astrophysics Supplement Series},
  volume    = {44},
  pages     = {363--370},
  year      = {1981}
}

@article{sofia_paper,
   title={SoFiA: a flexible source finder for 3D spectral line data},
   volume={448},
   ISSN={0035-8711},
   url={http://dx.doi.org/10.1093/mnras/stv079},
   DOI={10.1093/mnras/stv079},
   number={2},
   journal={Monthly Notices of the Royal Astronomical Society},
   publisher={Oxford University Press (OUP)},
   author = {{Serra, P.} and {Westmeier, T.} and {Giese, N.} and {Jurek, R.} and {Flöer, L.} and {Popping, A.} and {Winkel, B.} and {van der Hulst, T.} and {Meyer, M.} and {Koribalski, B. S.} and {Staveley-Smith, L.} and {Courtois, H.}},
   year={2015},
   month=feb, pages={1922-1929} }

@article{sofia_paper_2,
    author = {Westmeier, T and Kitaeff, S and Pallot, D and Serra, P and van der Hulst, J M and Jurek, R J and Elagali, A and For, B-Q and Kleiner, D and Koribalski, B S and Lee-Waddell, K and Mould, J R and Reynolds, T N and Rhee, J and Staveley-Smith, L},
    title = {sofia 2 - an automated, parallel H i source finding pipeline for the WALLABY survey},
    journal = {Monthly Notices of the Royal Astronomical Society},
    volume = {506},
    number = {3},
    pages = {3962-3976},
    year = {2021},
    month = {07},
    issn = {0035-8711},
    doi = {10.1093/mnras/stab1881},
    url = {https://doi.org/10.1093/mnras/stab1881},
    eprint = {https://academic.oup.com/mnras/article-pdf/506/3/3962/39458911/stab1881.pdf},
}

@INPROCEEDINGS{meerkat_paper,
       author = {{Jonas}, J. and {MeerKAT Team}},
        title = "{The MeerKAT Radio Telescope}",
    booktitle = {MeerKAT Science: On the Pathway to the SKA},
         year = 2016,
        month = jan,
          eid = {1},
        pages = {1},
          doi = {10.22323/1.277.0001},
       adsurl = {https://ui.adsabs.harvard.edu/abs/2016mks..confE...1J}
}

@article{askap_paper,
  author  = {Johnston, S. and Taylor, R. and Bailes, M. and others},
  title   = {Science with ASKAP},
  journal = {Experimental Astronomy},
  year    = {2008},
  volume  = {22},
  pages   = {151--273},
  doi     = {10.1007/s10686-008-9124-7}
}

@article{marcin2024,
    author = {Glowacki, M and Albrow, L and Reynolds, T and Elson, E and Mahony, E K and Allison, J R},
    title = {A serendipitous discovery of Hi-rich galaxy groups with MeerKAT},
    journal = {Monthly Notices of the Royal Astronomical Society},
    volume = {529},
    number = {4},
    pages = {3469-3483},
    year = {2024},
    month = {03},
    issn = {0035-8711},
    doi = {10.1093/mnras/stae684},
    url = {https://doi.org/10.1093/mnras/stae684},
    eprint = {https://academic.oup.com/mnras/article-pdf/529/4/3469/57074947/stae684\_supplemental\_file.pdf}
}

@article{Deg_2023,
  author = {Deg, N. and Palleske, R. and Spekkens, K. and Wang, J. and Jarrett, T. and English, J. and Lin, X. and Yeung, J. and Mould, J. R. and Catinella, B. and D{\'e}nes, H. and Elagali, A. and For, B.-Q. and Kamphuis, P. and Koribalski, B. S. and Lee-Waddell, K. and Murugeshan, C. and Oh, S. and Rhee, J. and Serra, P. and Westmeier, T. and Wong, O. I. and Bekki, K. and Bosma, A. and Carignan, C. and Holwerda, B. W. and Yu, N.},
  title = {WALLABY pilot survey: the potential polar ring galaxies NGC 4632 and NGC 6156},
  journal = {Monthly Notices of the Royal Astronomical Society},
  volume = {525},
  number = {3},
  pages = {4663--4684},
  year = {2023},
  month = sep,
  doi = {10.1093/mnras/stad2312},
  url = {http://dx.doi.org/10.1093/mnras/stad2312},
  issn = {1365-2966},
  publisher = {Oxford University Press (OUP)}
}

@inbook{haines-raytracing,
author = {Haines, Eric},
title = {Essential ray tracing algorithms},
year = {1989},
isbn = {0122861604},
publisher = {Academic Press Ltd.},
address = {GBR},
booktitle = {An Introduction to Ray Tracing},
chapter = {2},
pages = {33--77},
numpages = {45}
}

@misc{meta2025vrguidelines,
  author       = {{Meta Horizon OS Developers}},
  title        = {Guidelines for VR Performance Optimization},
  year         = {2025},
  howpublished = {\url{https://developers.meta.com/horizon/documentation/native/pc/dg-performance-guidelines/}},
  note         = {Accessed: 2025-07-18}
}

@phdthesis{sivitilli2023planetarium,
  author       = {Alexander Sivitilli},
  title        = {Characterizing the digital planetarium as a teaching and learning space},
  school       = {University of Cape Town},
  year         = {2023},
  type         = {PhD thesis},
  address      = {Cape Town, South Africa},
  url          = {http://hdl.handle.net/11427/40358},
  note         = {Supervised by Saalih Allie, Thomas Jarrett and Lucia Marchetti}
}

@article{taylor2025frelled,
  author  = {Rhys Taylor},
  title   = {FRELLED Reloaded: Multiple techniques for astronomical data visualisation in Blender},
  journal = {Astronomy and Computing},
  volume  = {51},
  pages   = {100927},
  year    = {2025},
  doi     = {10.1016/j.ascom.2024.100927}
}

@article{tudisco2025visivo,
  author       = {Giuseppe Tudisco and Fabio Vitello and Eva Sciacca and Ugo Becciani},
  title        = {From Local to Remote: VisIVO Visual Analytics in the Era of the Square Kilometre Array},
  journal      = {arXiv preprint},
  year         = {2025},
  eprint       = {2503.24113},
  archivePrefix= {arXiv},
  primaryClass = {astro-ph.IM},
  url          = {https://arxiv.org/abs/2503.24113}
}

@article{baddeley2000episodic,
  author  = {Alan D. Baddeley},
  title   = {The episodic buffer: a new component of working memory?},
  journal = {Trends in Cognitive Sciences},
  volume  = {4},
  number  = {11},
  pages   = {417--423},
  year    = {2000},
  doi     = {10.1016/S1364-6613(00)01538-2}
}

@misc{carta,
  author       = {Angus Comrie and
                  Kuo-Song Wang and
                  Yu-Hsuan Hwang and
                  Anthony Moraghan and
                  Pamela Harris and
                  Adrianna Pińska and
                  Carli Raul-Omar and
                  Cheng-Chin Chiang and
                  Ming-Yi Lin and
                  Tien-Hao Chang and
                  Rob Simmonds},
  title        = {CARTA: The Cube Analysis and Rendering Tool for
                   Astronomy
                  },
  month        = jan,
  year         = 2024,
  publisher    = {Zenodo},
  version      = {4.1.0},
  doi          = {10.5281/zenodo.15172686},
  url          = {https://doi.org/10.5281/zenodo.15172686},
}

@inproceedings{partiview,
  author    = {Levy, Steven and Chiang, Yi-Jen and Genetti, Jon and Larson, John and Reed, Daniel},
  title     = {Visualization for Astronomical Data Exploration},
  booktitle = {Astronomical Data Analysis Software and Systems X},
  editor    = {Harnden, F. R. and Primini, F. A. and Payne, H. E.},
  series    = {ASP Conference Series},
  volume    = {238},
  year      = {2001},
  pages     = {667}
}

@inproceedings{topcat,
  author    = {Taylor, M. B.},
  title     = {TOPCAT \& STIL: Starlink Table/VOTable Processing Software},
  booktitle = {Astronomical Data Analysis Software and Systems XIV},
  editor    = {Shopbell, P. and Britton, M. and Ebert, R.},
  series    = {ASP Conference Series},
  volume    = {347},
  year      = {2005},
  pages     = {29}
}

@article{slicerastro,
title = {SlicerAstro: A 3-D interactive visual analytics tool for HI data},
journal = {Astronomy and Computing},
volume = {19},
pages = {45-59},
year = {2017},
issn = {2213-1337},
doi = {https://doi.org/10.1016/j.ascom.2017.03.004},
url = {https://www.sciencedirect.com/science/article/pii/S2213133717300173},
author = {D. Punzo and J.M. van der Hulst and J.B.T.M. Roerdink and J.C. Fillion-Robin and L. Yu}
}

@misc{faherty2020,
  title         = {IDEAS: Immersive Dome Experiences for Accelerating Science},
  author        = {Jacqueline K. Faherty and Mark SubbaRao and Ryan Wyatt and Anders Ynnerman and Neil deGrasse Tyson and Aaron Geller and Maria Weber and Philip Rosenfield and Wolfgang Steffen and Gabriel Stoeckle and Daniel Weiskopf and Marcus Magnor and Peter K. G. Williams and Brian Abbott and Lucia Marchetti and Thomas Jarrrett and Jonathan Fay and Joshua Peek and Or Graur and Patrick Durrell and Derek Homeier and Heather Preston and Thomas Müller and Johanna M Vos and David Brown and Paige Giorla Godfrey and Emily Rice and Daniella Bardalez Gagliuffi and Alexander Bock and James Hedberg and Drew Rosen and Carter Emmart},
  year          = {2020},
  eprint        = {1907.05383},
  archiveprefix = {arXiv},
  primaryclass  = {astro-ph.IM},
  url           = {https://arxiv.org/abs/1907.05383}
}

@article{apertif_paper,
  title={Apertif: Phased array feeds for the westerbork synthesis radio telescope-system overview and performance characteristics},
  author={Van Cappellen, WA and Oosterloo, TA and Verheijen, MAW and Adams, EAK and Adebahr, B and Braun, R and Hess, KM and Holties, H and Van Der Hulst, JM and Hut, B and others},
  journal={Astronomy \& Astrophysics},
  volume={658},
  pages={A146},
  year={2022},
  publisher={EDP Sciences}
}

@article{wallaby_paper,
   title={WALLABY - an SKA Pathfinder H i survey},
   volume={365},
   ISSN={1572-946X},
   url={http://dx.doi.org/10.1007/s10509-020-03831-4},
   DOI={10.1007/s10509-020-03831-4},
   number={7},
   journal={Astrophysics and Space Science},
   publisher={Springer Science and Business Media LLC},
   author={Koribalski, Bärbel S. and Staveley-Smith, L. and Westmeier, T. and Serra, P. and Spekkens, K. and Wong, O. I. and Lee-Waddell, K. and Lagos, C. D. P. and Obreschkow, D. and Ryan-Weber, E. V. and Zwaan, M. and Kilborn, V. and Bekiaris, G. and Bekki, K. and Bigiel, F. and Boselli, A. and Bosma, A. and Catinella, B. and Chauhan, G. and Cluver, M. E. and Colless, M. and Courtois, H. M. and Crain, R. A. and de Blok, W. J. G. and Dénes, H. and Duffy, A. R. and Elagali, A. and Fluke, C. J. and For, B.-Q. and Heald, G. and Henning, P. A. and Hess, K. M. and Holwerda, B. W. and Howlett, C. and Jarrett, T. and Jones, D. H. and Jones, M. G. and Józsa, G. I. G. and Jurek, R. and Jütte, E. and Kamphuis, P. and Karachentsev, I. and Kerp, J. and Kleiner, D. and Kraan-Korteweg, R. C. and López-Sánchez, A. R. and Madrid, J. and Meyer, M. and Mould, J. and Murugeshan, C. and Norris, R. P. and Oh, S.-H. and Oosterloo, T. A. and Popping, A. and Putman, M. and Reynolds, T. N. and Rhee, J. and Robotham, A. S. G. and Ryder, S. and Schröder, A. C. and Shao, Li and Stevens, A. R. H. and Taylor, E. N. and van der Hulst, J. M. and Verdes-Montenegro, L. and Wakker, B. P. and Wang, J. and Whiting, M. and Winkel, B. and Wolf, C.},
   year={2020},
   month=jul }

@misc{VOTable_definition,
  author       = {Ochsenbein, F. and Williams, R. and Davenhall, C. and Demleitner, M. and Donaldson, T. and Durand, D. and Fernique, P. and Giaretta, D. and Hanisch, R. and McGlynn, T. and Szalay, A. and Taylor, M. and Wicenec, A.},
  title        = {VOTable Format Definition Version 1.5},
  year         = {2025},
  month        = jan,
  howpublished = {IVOA Recommendation},
  note         = {16 January 2025},
  url          = {https://ivoa.net/documents/VOTable/}
}

@ARTICLE{wcs_fits_paper,
       author = {{Calabretta}, M.~R. and {Greisen}, E.~W.},
        title = "{Representations of celestial coordinates in FITS}",
      journal = {A\&A},
         year = 2002,
        month = dec,
       volume = {395},
        pages = {1077-1122},
          doi = {10.1051/0004-6361:20021327},
archivePrefix = {arXiv},
       eprint = {astro-ph/0207413},
 primaryClass = {astro-ph},
       adsurl = {https://ui.adsabs.harvard.edu/abs/2002A&A...395.1077C}
}

@ARTICLE{vla_paper,
       author = {{Lacy}, M. and {Baum}, S.~A. and {Chandler}, C.~J. and {Chatterjee}, S. and {Clarke}, T.~E. and {Deustua}, S. and {English}, J. and {Farnes}, J. and {Gaensler}, B.~M. and {Gugliucci}, N. and {Hallinan}, G. and {Kent}, B.~R. and {Kimball}, A. and {Law}, C.~J. and {Lazio}, T.~J.~W. and {Marvil}, J. and {Mao}, S.~A. and {Medlin}, D. and {Mooley}, K. and {Murphy}, E.~J. and {Myers}, S. and {Osten}, R. and {Richards}, G.~T. and {Rosolowsky}, E. and {Rudnick}, L. and {Schinzel}, F. and {Sivakoff}, G.~R. and {Sjouwerman}, L.~O. and {Taylor}, R. and {White}, R.~L. and {Wrobel}, J. and {Andernach}, H. and {Beasley}, A.~J. and {Berger}, E. and {Bhatnager}, S. and {Birkinshaw}, M. and {Bower}, G.~C. and {Brandt}, W.~N. and {Brown}, S. and {Burke-Spolaor}, S. and {Butler}, B.~J. and {Comerford}, J. and {Demorest}, P.~B. and {Fu}, H. and {Giacintucci}, S. and {Golap}, K. and {G{\"u}th}, T. and {Hales}, C.~A. and {Hiriart}, R. and {Hodge}, J. and {Horesh}, A. and {Ivezi{\'c}}, {\v{Z}}. and {Jarvis}, M.~J. and {Kamble}, A. and {Kassim}, N. and {Liu}, X. and {Loinard}, L. and {Lyons}, D.~K. and {Masters}, J. and {Mezcua}, M. and {Moellenbrock}, G.~A. and {Mroczkowski}, T. and {Nyland}, K. and {O'Dea}, C.~P. and {O'Sullivan}, S.~P. and {Peters}, W.~M. and {Radford}, K. and {Rao}, U. and {Robnett}, J. and {Salcido}, J. and {Shen}, Y. and {Sobotka}, A. and {Witz}, S. and {Vaccari}, M. and {van Weeren}, R.~J. and {Vargas}, A. and {Williams}, P.~K.~G. and {Yoon}, I.},
        title = "{The Karl G. Jansky Very Large Array Sky Survey (VLASS). Science Case and Survey Design}",
      journal = {PASP},
         year = 2020,
        month = mar,
       volume = {132},
       number = {1009},
          eid = {035001},
        pages = {035001},
          doi = {10.1088/1538-3873/ab63eb},
archivePrefix = {arXiv},
       eprint = {1907.01981},
 primaryClass = {astro-ph.IM},
       adsurl = {https://ui.adsabs.harvard.edu/abs/2020PASP..132c5001L}
}

@article{jarrett2021exploring,
title = {Exploring and interrogating astrophysical data in virtual reality},
journal = {Astronomy and Computing},
volume = {37},
pages = {100502},
year = {2021},
issn = {2213-1337},
doi = {https://doi.org/10.1016/j.ascom.2021.100502},
url = {https://www.sciencedirect.com/science/article/pii/S2213133721000561},
author = {T.H. Jarrett and A. Comrie and L. Marchetti and A. Sivitilli and S. Macfarlane and F. Vitello and U. Becciani and A.R. Taylor and J.M. {van der Hulst} and P. Serra and N. Katz and M.E. Cluver}
}

@INPROCEEDINGS{marchetti_adass,
       author = {{Marchetti}, L. and {Jarrett}, T.~H. and {Comrie}, A. and {Sivitilli}, A.~K. and {McFarland}, S. and {Taylor}, A.~R. and {Clever}, M.},
        title = "{The DataToDome Initiative at the Iziko Planetarium in Cape Town and the IDIA Visualization Lab}",
    booktitle = {Astronomical Data Analysis Software and Systems XXIX},
         year = 2020,
       editor = {{Pizzo}, R. and {Deul}, E.~R. and {Mol}, J.~D. and {de Plaa}, J. and {Verkouter}, H.},
       series = {Astronomical Society of the Pacific Conference Series},
       volume = {527},
        month = jan,
        pages = {247},
       adsurl = {https://ui.adsabs.harvard.edu/abs/2020ASPC..527..247M}
}

@INPROCEEDINGS{sivitilli_adass,
       author = {{Sivitilli}, A.~K. and {Comrie}, A. and {Marchetti}, L. and {Jarrett}, T.~H.},
        title = "{Virtual Reality and Immersive Collaborative Environments: the New Frontier for Big Data Visualisation}",
    booktitle = {Astronomical Data Analysis Software and Systems XXIX},
         year = 2020,
       editor = {{Pizzo}, R. and {Deul}, E.~R. and {Mol}, J.~D. and {de Plaa}, J. and {Verkouter}, H.},
       series = {Astronomical Society of the Pacific Conference Series},
       volume = {527},
        month = jan,
        pages = {221},
          doi = {10.48550/arXiv.2103.14397},
archivePrefix = {arXiv},
       eprint = {2103.14397},
 primaryClass = {astro-ph.IM},
       adsurl = {https://ui.adsabs.harvard.edu/abs/2020ASPC..527..221S}
}

@inproceedings{Sivitilli2023iDaVIE,
  author       = {Alexander Sivitilli and Angus Comrie and Fabio Vitello and Lucia Marchetti and Thomas H. Jarrett and Thijs van der Hulst and Ug Becciani and Paulo Serra},
  title        = {An Update on iDaVIE (immersive Data Visualisation Interactive Explorer) and the Way Forward},
  booktitle    = {Astronomical Society of the Pacific Conference Series, Vol.\ 538, ADASS XXXII},
  pages        = {281},
  year         = {2023},
  doi          = {10.26624/LNCE9302},
  publisher    = {Astronomical Society of the Pacific}
}

@article{serra2023,
	author = {{Serra, P.} and {Maccagni, F. M.} and {Kleiner, D.} and {Molnár, D.} and {Ramatsoku, M.} and {Loni, A.} and {Loi, F.} and {de Blok, W. J. G.} and {Bryan, G. L.} and {Dettmar, R. J.} and {Frank, B. S.} and {van Gorkom, J. H.} and {Govoni, F.} and {Iodice, E.} and {Józsa, G. I. G.} and {Kamphuis, P.} and {Kraan-Korteweg, R.} and {Loubser, S. I.} and {Murgia, M.} and {Oosterloo, T. A.} and {Peletier, R.} and {Pisano, D. J.} and {Smith, M. W. L.} and {Trager, S. C.} and {Verheijen, M. A. W.}},
	title = {The MeerKAT Fornax Survey - I. Survey description and first evidence of ram pressure in the Fornax galaxy cluster},
	DOI= "10.1051/0004-6361/202346071",
	url= "https://doi.org/10.1051/0004-6361/202346071",
	journal = {A\&A},
	year = 2023,
	volume = 673,
	pages = "A146",
}

@misc{unity,
  author = {{Unity Technologies}},
  title = {Unity},
  year = {2021},
  note = {Version 2021.3.41f1 [Software]},
  howpublished = {\url{https://unity.com/}}
}

@misc{steamvr,
  author = {{Valve Corporation}},
  title = {SteamVR Unity Plugin},
  year = {2021},
  note = {Version 2.7.3 [Software]},
  howpublished = {\url{https://github.com/ValveSoftware/steamvr_unity_plugin}}
}

@article{ast_paper,
  title    = {AST: A library for modelling and manipulating coordinate systems},
  journal  = {Astronomy and Computing},
  volume   = {15},
  pages    = {33-49},
  year     = {2016},
  issn     = {2213-1337},
  doi      = {https://doi.org/10.1016/j.ascom.2016.02.003},
  url      = {https://www.sciencedirect.com/science/article/pii/S2213133716300129},
  author   = {David S. Berry and Rodney F. Warren-Smith and Tim Jenness}
}

@inproceedings{cfitsio_paper,
  author    = {{Pence}, William},
  title     = {{CFITSIO, v2.0: A New Full-Featured Data Interface}},
  booktitle = {Astronomical Data Analysis Software and Systems VIII},
  year      = 1999,
  editor    = {{Mehringer}, David M. and {Plante}, Raymond L. and {Roberts}, Douglas A.},
  series    = {Astronomical Society of the Pacific Conference Series},
  volume    = {172},
  month     = jan,
  pages     = {487},
  adsurl    = {https://ui.adsabs.harvard.edu/abs/1999ASPC..172..487P}
}

@article{hunter2007matplotlib,
  title={Matplotlib: A 2D graphics environment},
  author={Hunter, John D.},
  journal={Computing in Science \& Engineering},
  volume={9},
  number={3},
  pages={90--95},
  year={2007},
  publisher={IEEE},
  doi={10.1109/MCSE.2007.55}
}

@incollection{Marchetti2024,
author="Marchetti, Lucia
and Hulst, Thijs van der
and Jarrett, Thomas H.
and Comrie, Angus
and Sivitilli, Alexander
and Kirkham, Kechil",
editor="Vardoulaki, Eleni
and Dembska, Marta
and Drabent, Alexander
and Hoeft, Matthias",
title="Tools for Radio/Multi-Wavelength Big Data Interrogation",
bookTitle="Data-Intensive Radio Astronomy: Bringing Astrophysics to the Exabyte Era",
year="2024",
publisher="Springer International Publishing",
address="Cham",
pages="343--356",
isbn="978-3-031-58468-8",
doi="10.1007/978-3-031-58468-8_11",
url="https://doi.org/10.1007/978-3-031-58468-8_11"
}

@incollection{BADDELEY197447,
  title     = {Working Memory},
  author    = {Alan D. Baddeley and Graham Hitch},
  editor    = {Gordon H. Bower},
  booktitle = {Psychology of Learning and Motivation},
  series    = {Psychology of Learning and Motivation},
  volume    = {8},
  pages     = {47-89},
  year      = {1974},
  publisher = {Academic Press},
  issn      = {0079-7421},
  doi       = {https://doi.org/10.1016/S0079-7421(08)60452-1},
  url       = {https://www.sciencedirect.com/science/article/pii/S0079742108604521}
}

@INPROCEEDINGS{whisp_paper,
       author = {{van der Hulst}, J.~M. and {van Albada}, T.~S. and {Sancisi}, R.},
        title = "{The Westerbork HI Survey of Irregular and Spiral Galaxies, WHISP}",
    booktitle = {Gas and Galaxy Evolution},
         year = 2001,
       editor = {{Hibbard}, John E. and {Rupen}, Michael and {van Gorkom}, Jacqueline H.},
       series = {Astronomical Society of the Pacific Conference Series},
       volume = {240},
        month = jan,
        pages = {451},
       adsurl = {https://ui.adsabs.harvard.edu/abs/2001ASPC..240..451V}
}

@ARTICLE{Fornax_paper,
       author = {{Loni}, A. and {Serra}, P. and {Kleiner}, D. and {Cortese}, L. and {Catinella}, B. and {Koribalski}, B. and {Jarrett}, T.~H. and {Molnar}, D. Cs. and {Davis}, T.~A. and {Iodice}, E. and {Lee-Waddell}, K. and {Loi}, F. and {Maccagni}, F.~M. and {Peletier}, R. and {Popping}, A. and {Ramatsoku}, M. and {Smith}, M.~W.~L. and {Zabel}, N.},
        title = "{A blind ATCA HI survey of the Fornax galaxy cluster. Properties of the HI detections}",
      journal = {A\&A},
         year = 2021,
        month = apr,
       volume = {648},
          eid = {A31},
        pages = {A31},
          doi = {10.1051/0004-6361/202039803},
archivePrefix = {arXiv},
       eprint = {2102.01185},
 primaryClass = {astro-ph.GA},
       adsurl = {https://ui.adsabs.harvard.edu/abs/2021A&A...648A..31L}
}

@misc{idavie_zenodo,
  author    = {Jarrett, Thomas and
               Comrie, Angus and
               Sivitilli, Alexander and
               Pretorius, Pieter Cilliers and
               Vitello, Fabio and
               Marchetti, Lucia},
  title     = {{iDaVIE: Immersive Data Visualisation Interactive Explorer}},
  year      = {2024},
  month     = oct,
  publisher = {Zenodo},
  version   = {1.0},
  doi       = {10.5281/zenodo.13752029},
  url       = {https://doi.org/10.5281/zenodo.13752029},
  note      = {Software release},
}







\end{document}